\documentclass[11pt]{elsarticle}

\usepackage{lineno}
\usepackage[colorlinks=true,linkcolor=blue]{hyperref}

\modulolinenumbers[5]

\usepackage[margin=2.5cm]{geometry}% by courtesy of Mico
\makeatletter
\providecommand{\doi}[1]{%
	\begingroup
	\let\bibinfo\@secondoftwo
	\urlstyle{rm}%
	\href{http://dx.doi.org/#1}{%
		doi:\discretionary{}{}{}%
		\nolinkurl{#1}%
	}%
	\endgroup
}
\makeatother

\usepackage{graphicx}% Include figure files
\usepackage{dcolumn}% Align table columns on decimal point
\usepackage{bm}% bold math
\usepackage{hyperref}% add hypertext capabilities
\usepackage{epstopdf}
\usepackage{amsmath,esint}
\usepackage{amsmath}
\usepackage{amsfonts}
\usepackage{amssymb}
\usepackage{graphicx}
\usepackage{amsmath}
\usepackage{amsthm}
\usepackage{eucal}
\usepackage{amssymb}
\usepackage{mathrsfs}
\usepackage{float}
\usepackage{hyperref}
\usepackage{lineno}
\usepackage{multirow}
\usepackage{booktabs}
\usepackage{graphicx}
\usepackage{dashrule}
\usepackage{adjustbox}
\usepackage{xcolor}
\usepackage{graphicx}% Include figure files
\usepackage{dcolumn}% Align table columns on decimal point
\usepackage{bm}
\usepackage{natbib}
\usepackage{amssymb}
\usepackage{amsthm}
\usepackage{mathtools}
\usepackage{empheq}
\usepackage[most]{tcolorbox}
\usepackage{mathtools}
\usepackage{cleveref}
\usepackage{soul}
\usepackage{longtable}
\usepackage{amssymb}
\usepackage{mathrsfs}
\usepackage{amsmath,empheq}
\usepackage{multirow}
\usepackage{array}
\usepackage{amsthm}
\usepackage{booktabs}
\usepackage{amsmath,mathtools}
\usepackage{bm}% bold math
\usepackage{graphicx}
\usepackage{subcaption}

\usepackage{url}

\usepackage{tgpagella}
\usepackage{newpxmath}

\usepackage{caption}

\usepackage{tabularx}
\usetikzlibrary{datavisualization}

\usepackage{siunitx}

\usepackage{tikz}
\usetikzlibrary{shapes,arrows,chains}
\usetikzlibrary{calc}
\usetikzlibrary{positioning}

\usetikzlibrary{datavisualization}

\renewcommand{\vec}[1]{\boldsymbol{#1}}
\usepackage{xspace}
\let\storeBeta=\beta
\renewcommand\beta{\relax\ifmmode{\storeBeta}\else{$\storeBeta$}\fi\xspace}

\let\storeAlpha=\alpha
\renewcommand\alpha{\relax\ifmmode{\storeAlpha}\else{$\storeAlpha$}\fi\xspace}

\newcommand{\abs}[1]{\left| #1 \right|} % for absolute value
\renewcommand{\d}[2]{\frac{\mathrm{d} #1}{\mathrm{d} #2}} % for derivatives
\newcommand{\pd}[2]{\frac{\partial #1}{\partial #2}} 
\let\baraccent=\= % rename builtin command \= to \baraccent
\renewcommand{\=}[1]{\stackrel{#1}{=}} % for putting numbers above =

\newcount\colveccount
\newcommand*\colvec[1]{
	\global\colveccount#1
	\begin{pmatrix}
		\colvecnext
	}
	\def\colvecnext#1{
		#1
		\global\advance\colveccount-1
		\ifnum\colveccount>0
		\\
		\expandafter\colvecnext
		\else
	\end{pmatrix}
	\fi
}
\def\tphi{\widetilde{\phi}}
\def\tmu{\widetilde{\mu}}
\def\trhok{\widetilde{\rho}^{\, k}}
\def\tetak{\widetilde{\eta}^{\, k}}
\def\tp{\widetilde{p}}
\def\tvi{\widetilde{v}_i}

\def\tJi{\widetilde{J}_i}

\def\tpsi{\widetilde{\psi}}

\hypersetup{
	colorlinks,
	linkcolor={blue!50!black},
	citecolor={blue!50!black},
	urlcolor={blue!80!black},
	anchorcolor = {blue!80!black},
	filecolor = {blue!80!black},
	menucolor = {blue!80!black},
	runcolor = {blue!80!black}
}

\newtheorem{definition}{Definition}
\newtheorem{remark}{Remark}

\usepackage{pgfplots}
\pgfplotsset{
	table/search path={Figures},
}
	\usetikzlibrary{arrows,shapes}
	\pgfplotsset{compat=1.3}
	\pgfplotsset{
		discard if/.style 2 args={
			x filter/.code={
				\edef\tempa{\thisrow{#1}}
				\edef\tempb{#2}
				\ifx\tempa\tempb
				
				\fi
			}
		},
		discard if not/.style 2 args={
			x filter/.code={
				\edef\tempa{\thisrow{#1}}
				\edef\tempb{#2}
				\ifx\tempa\tempb
				\else
				
				\fi
			}
		}
	}
\usetikzlibrary{plotmarks}
\usetikzlibrary{spy}

\newcommand{\petsc}{\textsc{Petsc }}
\newcommand{\stampede}{\href{https://www.tacc.utexas.edu/systems/stampede2}{Stampede2 }}
\newcommand{\bridges}{\href{https://www.psc.edu/bridges}{Bridges}}
\usepackage{listings}
\usepackage{textcomp, gensymb}
\usepackage{xparse}

\NewDocumentCommand{\codeword}{v}{%
	\texttt{\textcolor{blue}{#1}}%
}

\definecolor{codegreen}{rgb}{0,0.6,0}
\definecolor{codegray}{rgb}{0.5,0.5,0.5}
\definecolor{codepurple}{rgb}{0.58,0,0.82}
\definecolor{backcolour}{rgb}{0.95,0.95,0.92}

\lstdefinestyle{mystyle}{
	backgroundcolor=\color{backcolour},   
	commentstyle=\color{codegreen},
	keywordstyle=\color{magenta},
	stringstyle=\color{codepurple},
	basicstyle=\ttfamily\footnotesize,
	breakatwhitespace=false,         
	breaklines=true,                 
	captionpos=b,                    
	keepspaces=true,                  
	showspaces=false,                
	showstringspaces=false,
	showtabs=false,                  
	tabsize=2
}

\usepackage{array}
\newcolumntype{C}[1]{>{\centering\arraybackslash}p{#1}}
\usepackage{xfp}
\usetikzlibrary{arrows,automata}

\allowdisplaybreaks

\usetikzlibrary{external}
\begin{document}

\begin{frontmatter}

%\title{Surface tension of molten metals using Rayleigh Theory of Oscillatin   g Force Free Droplets in a Two Phase Model}
\title{Modeling and simulations of high-density two-phase flows using projection-based Cahn-Hilliard Navier-Stokes equations}

\author[isuMechanicalEngAddress]{Ali Rabeh\fnref{equalContrib}}
\ead{arabeh@iastate.edu}
\author[isuMechanicalEngAddress,isuMathAddress]{Makrand A. Khanwale\fnref{equalContrib,MakFootnote}}
\ead{khanwale@stanford.edu}
%\author[isuMechanicalEngAddress]{Jonghyun~Lee}
\author[isuMechanicalEngAddress]{John J. Lee}
\ead{jolee@iastate.edu}

\author[isuMechanicalEngAddress]{Baskar~Ganapathysubramanian\corref{correspondingAuthor}}
\ead{baskarg@iastate.edu}

\cortext[correspondingAuthor]{Corresponding author}

\address[isuMechanicalEngAddress]{Department of Mechanical Engineering, Iowa State University, Iowa, USA 50011}
\address[isuMathAddress]{Department of Mathematics, Iowa State University, Iowa, USA 50011}
\fntext[equalContrib]{These authors contributed equally}
\fntext[MakFootnote]{Currently at the Department of Mechanical Engineering, Stanford University, CA, USA}
%\address{Department of Mechanical Engineering, Iowa State University, Iowa, USA 50011}

\begin{abstract}
Accurately modeling the dynamics of high-density ratio ($\mathcal{O}(10^5)$) two-phase flows is important for many material science and manufacturing applications. This work considers numerical simulations of molten metal oscillations in microgravity to analyze the interplay between surface tension and density ratio, a critical factor for terrestrial manufacturing applications. We present a projection-based computational framework for solving a thermodynamically-consistent Cahn-Hilliard Navier-Stokes equations for two-phase flows with large density ratios. The framework employs a modified version of the pressure-decoupled solver based on the Helmholtz-Hodge decomposition presented in Khanwale et al. [{\it A projection-based, semi-implicit time-stepping approach for the Cahn-Hilliard Navier-Stokes equations on adaptive octree meshes.}, Journal of Computational Physics 475 (2023): 111874]. We validate our numerical method on several canonical problems, including the capillary wave and single bubble rise problems. We also present a comprehensive convergence study to investigate the effect of mesh resolution, time-step, and interfacial thickness on droplet-shape oscillations. We further demonstrate the robustness of our framework by successfully simulating three distinct physical systems with extremely large density ratios ($10^4$–$10^5:1$), achieving results that have not been previously reported in the literature. 
\end{abstract}

\end{frontmatter}
\section{Introduction}
Simulating molten metal liquid drops suspended in microgravity is a classical problem in fluid mechanics studied by several researchers to understand their surface tension dynamics~\citep{Haas2021}. This interest stems from the fact that surface tension is an important parameter that can be observed in a wide range of practical applications such as in inkjet printing~\citep{Hoath2015}, additive manufacturing~\citep{YSLee2016}, and slug emulsification in continuous casting~\citep{Sumaria2017}. Understanding the oscillating behavior of these droplets contributes to optimizing industrial processes such as continuous casting~\citep{Hagemann2013} and holds promise for developing advanced materials with tailored properties. Accurate simulation of such phenomena supports industrial applications and advances fundamental understanding of fluid dynamics at extreme conditions, offering insights into the interplay of inertia, surface tension, and viscosity in regimes that are otherwise difficult to achieve experimentally. The measure of success in modeling and simulating suspended metal droplets is accurately reproducing the surface dynamics, thus capturing the underlying physics governing this behavior.

The Oscillating Droplet Method (ODM) is a standard microgravity experiment. In ODM, a molten droplet in micro (or zero) gravity is perturbed to undergo force-free small-amplitude axisymmetrical oscillations. These oscillations are directly related to the interplay between surface tension and viscous forces. The fluid surrounding the molten metal is a low-density gas, meaning that the force exerted by the gas can be neglected. Rayleigh's theory then allows one to analytically correlate the results of oscillation dynamics to the surface tension~\citep{Rayleigh1879}. Due to the difficulty and high cost of microgravity experiments, we are specifically interested in modeling the Oscillating Droplet Method. Modeling the ODM is challenging~\citep{Aalilija2020}, especially due to the very high density ratio and high viscosity ratios exhibited. These usually produce parasitic oscillations, and the stiff coupling often requires very small time steps, making this computationally very expensive.  
 
In this work, we explore and illustrate the capability of our recently developed framework for solving two-phase flows using a massively parallel adaptive mesh-based modeling framework. We deploy a thermodynamically consistent Cahn-Hilliard Navier-Stokes (CHNS)~\citep{Abels2012} model designed to satisfy the second law of thermodynamics to simulate our two-phase systems. The CHNS model is a diffuse interface model for the evolution of two-phase, immiscible, incompressible flows. An extensive amount of literature has been published on numerical approaches for solving the CHNS system for multiphase flows~\citep{Jacqmin1996, Jacqmin2000, Kim2004a, Feng2006, Shen2010a,Shen2010b, Han2015, Chen2016, Guo2017, Lowengrub1998, Khanwale2019, khanwale2023projection}. Unlike sharp interface models, the CHNS model uses a finite thickness transition region between the two phases to describe the interface between two immiscible fluids. The parameter controlling the interface thickness is then systematically reduced till the quantity of interest (oscillation dynamics) no longer changes, and we observe the sharp interface limit of the model. This numerical interfacial thickness is several orders of magnitude thicker than the actual physical interface, thus providing computational efficiency in modeling the oscillation dynamics. Such a diffuse-interface approach becomes especially critical in simulating high-density ratio systems, where traditional sharp interface models often face stability and accuracy challenges due to steep spatial gradients.

Simulating high density-ratio two-phase flows poses significant numerical challenges due to the steep gradients at the interface and the necessity of preserving both mass conservation and the solenoidality of the velocity field across all phases. To address these complexities, various approaches have been proposed in the literature. Diffuse interface methods, such as those based on the Cahn-Hilliard equation, provide an effective framework by representing sharp interfaces as smooth transition zones, enabling the handling of large density and viscosity contrasts up to 1000~\citep{Ding2007}. These methods are particularly advantageous for their ability to accurately resolve interfacial dynamics without compromising mass conservation. Complementing this, lattice Boltzmann models have demonstrated their robustness in capturing intricate interfacial phenomena, such as droplet splashing, with notable improvements in numerical stability and accuracy. By employing the Allen-Cahn phase-field equation, these models offer a simplified yet robust approach to high density-ratio flows~\citep{Liang2018}.

Here, for the first time to our knowledge, we use the CHNS model to simulate large density and viscosity ratio two-phase systems. The simulations of these extreme cases are mainly characterized by large spatial derivatives in the diffuse interface region, causing stiff non-linear systems. These stiffness challenges are further compounded by the need to ensure energy consistency and mass conservation, which are essential for accurately capturing the oscillatory behavior of droplets in microgravity. Our previous work~\citep{Khanwale2019} suggests that projection methods are a computationally attractive strategy for such stiff systems as they tend to be better at enforcing the solenoidality of the mixture velocity~\citep{Guermond2006}. Additionally, this gracefully allows the use of adaptive meshing by decoupling the pressure gradient from the momentum equation (and using the pressure Poisson equation to update the pressure field). Building on these insights, this work integrates the strengths of diffuse interface models and projection methods to achieve computational efficiency and physical accuracy in simulating extreme density and viscosity contrasts, while ensuring energy consistency and accurate resolution of interface dynamics. 

This work explores a few numerical issues as we simulate oscillating droplets with density ratio $\sim 10^5$. First, we explore and quantify the effect of interface mesh resolution. We find that using higher interfacial mesh resolution ensures energy decay without any instabilities, while lower mesh resolution causes the appearance of a small kink or instability early in the energy decay. This observation is especially critical because lower mesh resolution results violate the expected energy decay. Second, we also notice that extreme density-ratio cases require higher background mesh resolution. This high resolution becomes especially challenging for 3D simulations. This calls for adaptive mesh refinement to resolve the interface of our systems properly while still being computationally efficient. Specifically, the main contributions of this paper are as follows:
\begin{enumerate}
    \item \textbf{Effect of higher mesh resolution:} We conduct numerical experiments to investigate the impact of increased interfacial and background mesh resolution on the numerical method's energy stability, mass conservation, and stability, especially for simulations involving very large density ratios.
    \item \textbf{Scalable octree-based adaptive mesh refinement:} We implement a spatial discretization scheme using octree construction to improve the parallel performance of our framework. This approach enables efficient tracking of highly deforming interfaces through adaptive mesh refinement. Building on the method in \citep{khanwale2023projection}, we extend the framework to conduct large-scale 3D simulations at extreme density ratios.
    \item \textbf{Comprehensive test cases:} We validate the framework through a detailed series of test cases spanning a wide range of complexities, encompassing systems with both moderate and exceptionally high density ratios. These include scenarios such as capillary wave oscillations, single bubble rise dynamics, and simulations capturing intricate surface oscillations under microgravity conditions, showcasing the framework's robustness and adaptability across diverse physical regimes.
    \end{enumerate}

In addition to comparing our calculated interfacial tension from simulations with the experimental value, we verify through numerical experiments that our model is conditionally energy stable, providing the appropriate spatial and temporal resolution. We also verify numerically the mass conservation of the presented scheme. 
\section{Governing equations}
\label{sec:govern_equations}

We use the thermodynamically consistent model of Cahn-Hilliard Navier-Stokes equations from ~\citet{Abels2012}.  We consider a bounded domain $\Omega \subset \mathbb{R}^3$, containing the two fluids (metal droplet and surrounding fluid) and a time interval, $[0, T]$. 
We define a phase field, $\phi$, that tracks the fluids, i.e. takes a value of $+1$, and $-1$ in domains occupied by the surrounding fluid and the metal droplet, respectively.
%Let $\rho_{+}$ ($\eta_{+}$ ) and $\rho_{-}$ ($\eta_{-}$) denote the specific density (viscosity) of the metal droplet and the surrounding gas, respectively.  
Let $\rho_{+}$ and $\rho_{-}$ be the specific densities of the heavier fluid and lighter fluid respectively. Similarly, let $\eta_{+}$ and $\eta_{-}$ be the dynamic viscosities of the heavier fluid and lighter fluid respectively. Then for a smoothly varying phase field, the non-dimensional mixture density\footnote{Our non-dimensional form uses the specific density/viscosity of the heavier fluid i.e. $\rho_{+}$ ($\eta_{+}$) as the scaling density (viscosity).} is given by $\rho(\phi) = \alpha\phi + \beta$, where $\alpha = \frac{\rho_{-} - \;\rho_{+}}{2\rho_{+}}$ and $\beta = \frac{\rho_{-} + \;\rho_{+}}{2\rho_{+}}$. Similarly, non-dimensional mixture viscosity is given by $\eta(\phi) = \gamma\phi + \xi$, where $\gamma = \frac{\eta_{-} - \;\eta_{+}}{2\eta_{+}}$ and $\xi = \frac{\eta_{-} + \;\eta_{+}}{2\eta_{+}}$. 
Following the non-dimensionalization (as detailed in~\citep{khanwale2021energy}), the non-dimensional governing equations are given by:
\begin{align}
\begin{split}
\text{Momentum Eqns:} & \quad \pd{\left(\rho(\phi) v_i\right)}{t} + \pd{\left(\rho(\phi)v_iv_j\right)}{x_j} + \frac{1}{Pe}\pd{\left(J_jv_i\right)}{x_j} +\frac{Cn}{We} \pd{}{x_j}\left({\pd{\phi}{x_i}\pd{\phi}{x_j}}\right) \\
& \quad \quad \quad + 
\frac{1}{We}\pd{p}{x_i} - \frac{1}{Re}\pd{}{x_j}\left({\eta(\phi)\pd{v_i}{x_j}}\right) - \frac{\rho(\phi)\hat{{g_i}}}{Fr} = 0,
\label{eqn:nav_stokes} 
\end{split} \\
\text{Thermo Consistency:} & \quad J_i = \frac{\left(\rho_- - \rho_+\right)}{2 \rho_+Cn} \, m(\phi) 
\, \pd{\mu}{x_i}, \label{eqn:thermoConsistecyJTerm}\\
\text{Solenoidality:} & \quad \pd{v_i}{x_i} = 0, \label{eqn:solenoidality}\\
\text{Continuity:} & \quad \pd{\rho(\phi)}{t} + \pd{\left(\rho(\phi)v_i\right)}{x_i}+
\frac{1}{Pe} \pd{J_i}{x_i} = 0, \label{eqn:cont}\\
\text{Chemical Potential:} & \quad \mu = \psi'(\phi) - Cn^2 \pd{}{x_i}\left({\pd{\phi}{x_i}}\right) ,\label{eqn:mu_eqn} 
\\ 
\text{Cahn-Hilliard Eqn:} & \quad \pd{\phi}{t} + \pd{\left(v_i \phi\right)}{x_i} - \frac{1}{PeCn} \pd{}{x_i}\left({m(\phi)\pd{\mu}{x_i}}\right) = 0. 
\label{eqn:phi_eqn}
\end{align}
In the above equations, $\vec{v}$ is the \textit{volume fraction weighted mixture velocity}\footnote{We use Einstein notation throughout the manuscript. In this notation, $v_i$ represents the $i^{\text{th}}$ component of the vector $\vec{v}$, and any repeated index is implicitly summed over.}, $p$ is the volume averaged pressure, $\phi$ is the phase field (interface tracking variable), and $\mu$ is the chemical potential. Mobility $m(\phi)$ is assumed to be a constant with a value of one. We use the Ginzburg-Landau polynomial form of the free energy density defined as follows,
\begin{align}
\psi(\phi) = \frac{1}{4}\left( \phi^2 - 1 \right)^2 \quad \text{and} \qquad
\psi'(\phi) = \phi^3 - \phi.
\end{align}

The non-dimensional parameters are defined as, Reynolds, $Re = \frac{\rho_{+} u_{r} D}{\eta_{+}}$, Weber, $We = \frac{\rho_{+}u_{r}^2 D}{\sigma}$, Cahn, $Cn = \frac{\varepsilon}{D}$, Peclet, $Pe = \frac{u_{r} D^2}{m_{r}\sigma}$, and Froude, $Fr = \frac{u_{r}^2}{g D}$. In the above non-dimensionalization, $u_{r}$ and $D$ denote the reference velocity and length scale, respectively. Here, $m_{r}$ is the reference mobility, $\sigma$ is the scaling interfacial tension, $\varepsilon$ is the interface thickness. %In our case, we assume that the gravitational effects are negligible because the corresponding experiments are conducted in microgravity. 

\citet{Abels2012} showed that the system of equations \cref{eqn:nav_stokes} -- \cref{eqn:phi_eqn} has a dissipative law given by, 
\begin{equation}
\d{E_{\mathrm{tot}}}{t} = -\frac{1}{Re} \int_{\Omega} \frac{\eta(\phi)}{2} \sum_{i} \sum_{j} \left| \frac{\partial v_i}{\partial x_j} \right|^2 \mathrm{d}\vec{x} - \frac{Cn}{We} \int_{\Omega} \sum_{i} m(\phi) \left| \frac{\partial \mu}{\partial x_i} \right|^2 \mathrm{d}\vec{x},
\end{equation}
where the total energy is given by,  
\begin{equation}
E_{\mathrm{tot}}(\vec{v},\phi, t) = \int_{\Omega}\frac{1}{2}\rho\left(\phi\right) \sum_{i} \left| v_i \right|^2 \mathrm{d}\vec{x} + \frac{1}{CnWe}\int_{\Omega} \left(\psi(\phi) + \frac{Cn^2}{2} \sum_{i} \left| \frac{\partial \phi}{\partial x_i} \right|^2 + \frac{1}{Fr} \rho(\phi) y \right) \mathrm{d}\vec{x}.
\label{eqn:energy_functional}
\end{equation}
%The norms used in the above expression are the Euclidean vector norm and the Frobenius matrix norm:

%\begin{equation}
%\norm{\vec{v}}^2 := \sum_i \abs{v_i}^2 \qquad \text{and} \qquad 
%\norm{\nabla\vec{v}}^2_F := \sum_i \sum_j \abs{\frac{\partial v_i}{\partial x_j}}^2.
%\end{equation}
The energy dissipation law above is valid for closed systems where there are no energy fluxes through the boundaries. Our choice of boundary conditions for the simulations conform to this requirement and we do not have to modify the above energy dissipation law for boundary effects. 

%\subsection{Initial conditions, boundary conditions, and non-dimensionalization}

\section{Numerical method}
\label{sec:numerical_method}
We employ the semi-implicit time-marching scheme in \citet{khanwale2023projection} for solving the Navier-Stokes equations, incorporating a projection step to separate the pressure calculation. This approach uses a fully implicit time-integration scheme for the Cahn-Hilliard equations and a semi-implicit time-integration scheme for momentum equations. An incremental pressure-projection scheme that preserves energy stability is used. The numerical method is included here for completeness.

Let $\delta t$ be a constant time-step, and time (uniformly spaced) $t^k := k \delta t$; we define the following time-averages: 
\begin{align}
\begin{split}
& \quad \widetilde{\vec{v}}^{k} := \frac{\vec{u}^{k} + \vec{v}^{k+1}}{2}, \quad \widetilde{\vec{u}}^{k} := \frac{\vec{u}^{k} + \vec{u}^{k+1}}{2}, \quad \widehat{\vec{u}}^{k} := \frac{3\vec{u}^{k} + \vec{u}^{k-1}}{2}, \quad \tp^{k} := \frac{{p}^{k+1}+{p}^{k}}{2}, \quad 
\tphi^{k} := \frac{{\phi}^{k+1} + {\phi}^{k}}{2}, \\ & \quad
\text{and} \quad \tmu^{k} := \frac{{\mu}^{k+1} + {\mu}^{k}}{2},
\end{split}
\end{align}

and the following potential function evaluations:

\begin{align}
\begin{split}
	\label{eqn:psi_ave_def}
	& \quad \tpsi^k := \psi\left( \tphi^k \right), \quad 
	\tpsi'^{k} := \psi'\left( \tphi^k \right), \quad
	\trhok := \rho\left(\tphi^k\right), \quad 
	\rho^{k+1} := \rho\left(\phi^{k+1}\right), \quad 
	\rho^{k} := \rho\left(\phi^{k}\right), \\ & \quad
	\text{and} \quad
	\tetak := \eta\left(\tphi^k\right).
	\end{split}
\end{align}
It is important to note that two distinct velocity fields are involved in this method. The intermediate velocity, denoted as $\widetilde{\vec{v}}^{k}$, is computed first, while the end-of-step solenoidal velocity, $\widetilde{\vec{u}}^{k}$, satisfies the incompressibility constraint. Additionally, $\widehat{\vec{u}}^{k}$ represents an explicitly averaged solenoidal velocity, calculated using values from previous timesteps. With these definitions, the discrete in time continuous in space method is based on the variational form of Cahn-Hilliard Navier-Stokes (CHNS) equations in~\citep{khanwale2023projection}, which we define below. We do not use the conventional superscript $h$, typically used to denote finite-dimensional conforming Galerkin approximations, even though the function spaces in the definition below are finite-dimensional.

\begin{definition}
	Let $(\cdot,\cdot)$ be the standard $L^2$ inner product. The time-discretized variational problem can be stated as follows: find $\vec{v}^{k+1}(\vec{x}) \in \vec{H}_0^1(\Omega)$, $p^{k+1}(\vec{x})$, $\phi^{k+1}(\vec{x})$, $\mu^{k+1}(\vec{x})$ $\in {H}^1(\Omega)$ such that
	\begin{align}
	\begin{split}
		\text{Velocity Prediction:} & \quad \left(w_i, \, \trhok \, \frac{v^{k+1}_i - u^k_i}{\delta t}\right) + \left(w_i, \trhok \widehat{u_j}^{k} \, \pd{\tvi^{k}}{x_j}\right) \\ & \quad +
		\frac{1}{Pe}\left(w_i, \, \widehat{J_j}^{k} \, \pd{\tvi^{k}}{x_j}\right) -
		\frac{Cn}{We} \left(\pd{w_i}{x_j}, \, {\pd{\tphi^{k}}{x_i}\pd{\tphi^{k}}{x_j}}\right) + \frac{1}{We}\left(w_i, \, \pd{p^{k}}{x_i} \right) \\ & \quad + \frac{1}{Re}\left(\pd{w_i}{x_j}, \,\tetak {\pd{\tvi^{k}}{x_j}} \right) -
		\left(w_i,\frac{\trhok \, \widehat{g_i}}{Fr}\right) = 0, 
		\label{eqn:nav_stokes_var_semi_disc}
	\end{split} \\ 
	\text{Projection:} & \quad \left(w_i, \trhok \frac{u_i^{k+1} - v_i^{k+1}}{\delta t} \right) +
	\frac{1}{2We} \left(w_i, \pd{ \left(p^{k+1} - p^{k}\right)}{x_i} \right) = 0,
	 \label{eqn:projection_semi_disc} \\
	\text{Continuity:} & \quad \left(q, \frac{\rho^{k+1} - \rho^k}{\delta t} \right) +
	\left(q, \pd{\left(\trhok \widehat{u_j}^{k}\right)}{x_i} \right) - \frac{1}{Pe} \left(\widehat{J_j}^{k}, {\pd{q}{x_j}} \right) = 0,
	 \label{eqn:continuity_semi_disc} \\
	\text{Thermo Consistency:} & \quad \tJi^{k} = \frac{\left(\rho_- - \rho_+ \right)}{2 \rho_{+}Cn} \, \pd{\tmu^{k}}{x_i}, \label{eqn:thermo_consistency_semi_disc} \\
	\text{Solenoidality:} & \quad 
	%\left(q, \, \pd{v^{k}_{i}}{x_i}\right) = 0, 
	\quad \left(q, \, \pd{u^{k+1}_{i}}{x_i}\right) = 0,
	\label{eqn:solenoidality_var_semi_disc} \\
	\text{Chemical Potential:} & \quad 
	-\left(q,\tmu^{k}\right) + \left(q, \tpsi'^{k} \right) + Cn^2 \left(\pd{q}{x_i}, \, {\pd{\tphi^{k}}{x_i}}\right) = 0, \label{eqn:mu_eqn_var_semi_disc}\\
	\text{Cahn-Hilliard Eqn:} & \quad \left(q, \frac{\phi^{k+1} - \phi^k}{\delta t} \right) -
	\left(\pd{q}{x_i}, \, \widetilde{u_i}^{k} \tphi^{k} \right) + \frac{1}{PeCn} \left(\pd{q}{x_i}, \, {\pd{\tmu^{k}}{x_i}} \right) = 0,
	\label{eqn:phi_eqn_var_semi_disc}
	\end{align}
	$\forall \vec{w} \in \vec{H}^1_0(\Omega)$, $\forall q \in H^1(\Omega)$, given $\vec{u}^{k},\vec{u}^{k-1} \in \vec{H}_0^1(\Omega)$, and $\phi^{k},\mu^{k} \in H^1(\Omega)$.
	\label{def:variational_form_sem_disc}

\end{definition} 

We redefine the pressure by absorbing the coefficient $\frac{1}{We}$ in its definition. Using the solenoidality of $u^{k+1}$, \cref{eqn:projection_semi_disc} is implemented in a two-step strategy.

\begin{definition}
	Let $(\cdot,\cdot)$ be the standard $L^2$ inner product. The time-discretized variational problem can be stated as follows: find $\vec{u}^{k+1}(\vec{x}) \in \vec{H}_0^1(\Omega)$, $p^{k+1}(\vec{x})$ $\in {H}^1(\Omega)$ such that
\begin{align}
 \text{Pressure Poisson:} & \quad \left(\pd{q}{x_i}, \frac{1}{\trhok} \pd{p^{k+1}}{x_i} \right)=\frac{-2}{ \delta t} \left(q, \pd{v_i^{k+1}}{x_i}\right) +\left(\pd{q}{x_i}, \frac{1}{\trhok} \pd{p^{k}}{x_i} \right) \label{eqn:pressure_poisson_var_semi_disc} \\
\text{Velocity update:} & \quad \left(w_i, \trhok u_i^{k+1} \right) + \frac{\delta t}{2}\left(w_i, \pd{p^{k+1}}{x_i} \right)=\left(w_i, \trhok v_i^{k+1} \right) + \frac{\delta t}{2}\left(w_i, \pd{p^{k}}{x_i} \right) \label{eqn:velocity_update_var_semi_disc}
\end{align}
$\forall \vec{w} \in \vec{H}^1_0(\Omega)$, given $\vec{v}^{k+1} \in \vec{H}_0^1(\Omega)$, and $p^{k} \in H^1(\Omega)$.
\label{def:pp_velocityUpdate_sem_disc}
\end{definition}
 
Pressure (\cref{eqn:pressure_poisson_var_semi_disc}) is solved by dividing the strong form corresponding to \cref{eqn:projection_semi_disc} by $\trhok$ and then taking the divergence of the equation and weakening the result subsequently. Velocity is then updated using the current pressure by using the Helmholtz-Hodge decomposition equation \cref{eqn:velocity_update_var_semi_disc}.

%\begin{remark}
 In \Cref{def:variational_form_sem_disc} and \Cref{def:pp_velocityUpdate_sem_disc} for the momentum and projection equations, the boundary terms in the variational form are zero because the velocity and the basis functions live in $\vec{H}_0^1(\Omega)$. Also we use the no flux boundary condition for $\phi$ and $\mu$, which makes the boundary terms zero (i.e., $\left(q,\pd{\tphi^k}{x_i} \hat{n_i} \right)$ = 0 and $\left(q,\pd{\mu^k}{x_i} \hat{n_i} \right)$ = 0). This choice of boundary conditions ensures the disappearance of boundary terms when the equations are weakened in the variational form. 
 %All examples in this paper use these types of boundary conditions.
%\end{remark}

\begin{remark} While $\phi \in [-1, 1]$ in the original equations, there is a possibility of bound violation of $\phi$ due to numerical errors. While this does not adversely affect the $\phi$ evolution (i.e., the CH equation), it may cause non-positivity of the mixture density $\rho(\phi)$ and the mixture viscosity $\eta(\phi)$, which directly depend on $\phi$. This causes a drift of the bulk phase density from the true specific density of that phase, with some locations exhibiting negative density (or viscosity). This effect is especially possible for a high-density ratio between the two fluids considered in this paper. A simple fix for this issue is saturation scaling, i.e., pulling back the value of $\phi$ only for calculating the density and viscosity. We, therefore, define $\phi^s$ that is only used for the calculation of mixture density and viscosity, where $\phi^s$ is given by: 
	\begin{equation}
	\phi^s := 
	\begin{cases}
	\phi, &\quad \text{if} \;\; \abs{\phi} \leq 1, \\
	\mathrm{sign}(\phi), &\quad \text{otherwise.} 
	\end{cases}
	\label{eqn:phi_for_density}
	\end{equation}
\label{rmk:phi_pullback} 
\end{remark}
\begin{remark}
	\citet{khanwale2023projection} showed that the numerical method in~\cref{def:variational_form_sem_disc} is essentially unconditionally energy stable. This implies that the difference between $E_{\text{tot}}(\vec{v},\phi, t)$ (see \cref{eqn:energy_functional}) calculated between two successive time-steps is always negative. We also numerically verify another important property of the time-scheme in~\cref{def:variational_form_sem_disc}, which is mass conservation.
	\label{rmk:energy_mass} 
\end{remark}

We solve the spatially discretized version of variational problems in \Cref{def:variational_form_sem_disc} and \Cref{def:pp_velocityUpdate_sem_disc}
using a block iteration technique, i.e., we treat the velocity prediction equations (\cref{eqn:nav_stokes_var_semi_disc}), pressure
Poisson equation (\cref{eqn:pressure_poisson_var_semi_disc}), velocity update equations (\cref{eqn:velocity_update_var_semi_disc}), and the Cahn-Hilliard equations (\cref{eqn:mu_eqn_var_semi_disc} to \cref{eqn:phi_eqn_var_semi_disc}) as distinct sub-problems. Thus, three linear solvers of (1) velocity prediction, (2) pressure Poisson, and (3) velocity update are stacked together with a non-linear solver for Cahn-Hilliard equations inside the time loop. These solvers are each solved twice (two blocks) within every time step to preserve the order of accuracy and self-consistency. We solve the equations using a block iterative approach, allowing us to make the coupling
variables from one equation constant in the other during each respective linear/non-linear solve. We adopt a strategy here such that the mixture velocity used in the advection of the phase field (\cref{eqn:phi_eqn_var_semi_disc}) is always weakly solenoidal. This means that the pressure Poisson and velocity update is performed two times as we run the block iteration twice. This strategy is robust, especially for our high density ratios where the pressure gradients between the two phases can be high. The detailed framework flowchart and the octree-based domain decomposition can be found in~\citep{khanwale2023projection}.

\subsection{Variational multiscale stabilization}
\label{subsec:scaling_tauM}

We discretize the unknowns in space using conforming continuous Galerkin (cG) finite elements with piecewise polynomial approximations. However, using the same polynomial order for both velocity and pressure can cause numerical instabilities, as it violates the discrete inf-sup condition \cite{John2002}. Although the projection method separates the pressure and velocity, employing equal-order basis functions for both still results in instabilities. To address this issue, we apply the residual-based variational multi-scale (VMS) method \citet{Bazilevs2007}, which introduces stabilization terms to the formulation and circumvents the inf-sup stability condition. 

If $\vec{v} \in \vec{V}$ and $p \in Q$, then we decompose these spaces as: 
\[
\vec{V} = \vec{V}^c \oplus \vec{V}^f \quad \text{and} \quad Q = Q^c \oplus Q^f,
\]
where $\vec{V}^c$ and $Q^c$ are the finite-dimensional cG(1) subspaces of $\vec{V}$ and $Q$, respectively, and the superscript $f$ versions are the complements of the cG(1) subspaces in $\vec{V}$ and $Q$, respectively.

We use the residual-based approximation proposed in \citet{Bazilevs2007} by decomposing the velocity as follows: 
\begin{align}
\vec{v}= \vec{v}^c+\vec{v}^f,
\end{align}
where $\vec{v}^c$ is the coarse-scale velocity and $\vec{v}^f$ is the fine-scale velocity. 
We define a projection operator, $\mathscr{P}:\vec{V} \rightarrow \vec{V}^{c}$, such that
 $\vec{v}^{c} = \mathscr{P}\{\vec{v}\}$ and $\vec{v}^{f} = \vec{v} - \mathscr{P}\{\vec{v}\}$. Note that the $\vec{V}^{c}$ is a finite-dimensional subspace of $\vec{V}$. We denote our finite dimensional subspaces with a superscript $r$. For example, $\vec{v}^{c}(\vec{x}) \in \mathscr{P}\vec{H}_0^{1,r}(\Omega)$. In our case, $\vec{V}^{c}$ is also our finite element space. 

The fine-scale components, applied to a two-phase system, to close the equations:
\begin{align}
 \trhok v_i^{f,k+1} = -\tau_m \mathcal{R}_m \left(\trhok,\vec{v}^{k+1},\vec{u}^k,\vec{u}^{k-1}, p^k \right)
\end{align}
with the following stabilization parameter:
\begin{align}
\tau_m &= \left(\frac{4}{\delta t^2} + {v_{\text{eff}_i}}^cG_{ij}{v_{\text{eff}_j}}^c + C_{I} \left(\frac{\tetak}{\trhok Re}\right)^2 G_{ij}G_{ij}\right)^{-1/2}\label{eqn:tau_m}\\ 
G_{ij} &= \sum_{k=1}^{d} \frac{\partial \xi_k}{\partial x_i} \frac{\partial \xi_k}{\partial x_j} 
\end{align}

We set $C_{I}$ to 6 as we restrict our simulations to linear basis functions. $G_{ij}$ is a mesh tensor that accounts for the mapping between the parametric and physical domain of the element. For a uniform mesh of equal aspect ratio $(\Delta x = \Delta y = h)$, $G_{ij}=\left(\frac{2}{h}\right)^2$. The term ${v_{\text{eff}_j}}^c$ represents the effective convective velocity, which combines contributions from both convective and diffusive fluxes in the momentum equation~\cref{eqn:nav_stokes_var_semi_disc}:

\begin{align}
 \left(w_i, \left(\rho u_j + \frac{J_j}{Pe}\right) \frac{dv_i}{dx_j}\right)  
\end{align}

The effective velocity in~\cref{eqn:tau_m} is then defined as:
\begin{align}
 v_{\text{eff}_j} = u_j + \frac{J_j}{\rho Pe}  
\end{align}

The full expression for the stabilization parameter then becomes:

\begin{align}
\tau_m &= \left(\frac{4}{\delta t^2} + \left(u_i^c + \frac{\widehat{J_i}}{\rho Pe}\right) G_{ij} \left(u_j^c + \frac{\widehat{J_j}}{\rho Pe} \right) + C_{I} \left(\frac{\tetak}{\trhok Re}\right)^2 G_{ij}G_{ij}\right)^{-1/2}\label{eqn:tau2_m}
\end{align}

This stabilization parameter accounts for time-stepping constraints, mesh properties, and effective velocity contributions from convective and diffusive terms (different from $\tau_m$ in ~\citep{khanwale2023projection}). The momentum residual is also given by 

\begin{align}
\begin{split}
\mathcal{R}_m \left(\trhok , v_i^{c,k+1}, u_i^k, u_i^{k-1}, p^k \right) = \trhok \frac{v_i^{c,k+1}-u_i^k}{\delta t} + \trhok \widehat{u_j}^{k} \pd{\tvi^{c,k}}{x_j} +\frac{1}{Pe} \widehat{J_j} \pd{\tvi^{c,k}}{x_j} \\
+\frac{Cn}{We}\pd{}{x_j}\left({\pd{\phi^{h,k}}{x_i}\pd{\phi^{h,k}}{x_j}} \right) + \frac{1}{We} \pd{p^k}{x_i} -\frac{1}{Re}\pd{}{x_j} \left(\tetak\pd{\tilde{v_i}^{c,k}}{x_j} \right) - \frac{\trhok \hat{g_i}}{Fr}\label{eqn:R_m}\\ 
\end{split}
\end{align}

The use of the intermediate velocity ($v_i^{c,k+1}$) instead of ($u_i^{c,k+1}$) in the momentum residual introduces a consistency error. However, this step is unavoidable, as only $v_i^{c,k+1}$ is available at the intermediate step. While the intermediate velocity is not strictly divergence-free, it can be considered semi-solenoidal and approximates $u_i^{c,k+1}$ in the residual equation. Importantly, this consistency error diminishes and approaches zero as the timestep size decreases.

%Move to s-tuning paper: Our formulation uses the continuous Galerkin (cG) method, which results in poor solenoidality, particularly near the interface of high density ratio two-phase flows. Since the fine-scale velocity is inversely proportional to density, $\frac{1}{\rho}$, this results in significant variations in the term $\left(q, \pd{v_i}{x_i}\right)$, causing numerical instabilities as the velocity and pressure blow up in the low density phase. To mitigate these instabilities, we reduce the stabilization parameter, $\tau_m$, by multiplying it with a constant scaling factor $s$. This parameter was determined empirically, and we numerically tested various values to identify the lowest value that produced stable results with good solenoidality. Note that this modification is only needed for high density ratio flows. This approach is inspired by similar strategies employed in the context of thin structures in \citet{Kamensky2015}, where they weakened the stabilization near immersed shell structures to address issues with compressibility arising from large pressure gradients. 

Let $\Omega \subset \mathbb{R}^d$ be the spatial domain, and let $\mathcal{T}_h$ be a conforming finite element discretization of $\Omega$. The $L^2$ inner product $(\cdot, \cdot)$ is defined as:
\[ (u, v) = \int_\Omega u \, v \, d\Omega \approx \sum_{K \in \mathcal{T}_h} \int_K u \, v \, dK, \]
where $\int_K$ denotes the integration over each finite element $K$, the finite-dimensional test and trial functions are chosen from the finite element subspaces $\mathscr{P}\vec{H}_0^{1,r}(\Omega)$ and $\mathscr{P}{H}^{1,r}(\Omega)$, which consist of piecewise polynomial functions defined on $\mathcal{T}_h$. The stabilization terms, such as those involving $\tau_m R_m$, are also projected onto these finite-dimensional spaces. The residual terms $\mathcal{R}_m$ in the VMS formulation involve fine-scale corrections and stabilization that are inherently evaluated in the finite element inner product context. These residual terms are not directly evaluated in $L^2$, but rather in the finite element inner product space, which approximates $L^2$ integrals over the discrete domain. This distinction ensures that all terms are consistent with the finite element framework.

\begin{definition}
Let $(\cdot,\cdot)$ be the finite element $L^2$ inner product defined on the discretized domain. The time-discretized variational problem using variational multiscale stabilization can be stated as follows: find $\vec{v}^{c,k+1}(\vec{x}) \in \mathscr{P}\vec{H}_0^{1,r}(\Omega)$, $p^{k+1}(\vec{x})$, $\phi^{k+1}(\vec{x})$, $\mu^{k+1}(\vec{x})$ $\in \mathscr{P}{H}^{1,r}(\Omega)$ such that
\begin{align}
\begin{split}
		\text{Velocity Prediction:} & \quad \left(w_i, \, \trhok \, \frac{v^{c,k+1}_i - u^k_i}{\delta t}\right) + \frac{1}{2} \left(w_i, \trhok \widehat{u_j}^{k} \, \pd{v^{c,k+1}_i}{x_j}\right) + \frac{1}{2} \left(w_i, \trhok \widehat{u_j}^{k} \, \pd{u_i^{k}}{x_j}\right) \\ 
  & + \frac{1}{2Pe}\left(w_i, \, \widehat{J_j}^{k} \, \pd{v^{c,k+1}_i}{x_j}\right) + \frac{1}{2Pe}\left(w_i, \, \widehat{J_j}^{k} \, \pd{u_i^{k}}{x_j}\right) \\
  & \boxed{+ \frac{1}{2}\left(\pd{w_i}{x_j}, \, \left(\widehat{u_j}^{k} + \frac{1}{Pe \trhok} \widehat{J_j}^{k} \right) \, \tau_m R_m(\trhok,\Tilde{\eta_k}, v_i^{c,k+1}, u_i^k, \phi^k, p^k) \right)}   \\
  & - \frac{Cn}{We} \left(\pd{w_i}{x_j}, \, {\pd{\tphi^{h,k}}{x_i}\pd{\tphi^{h,k}}{x_j}}\right) + \frac{1}{We}\left(w_i, \, \pd{p^{k}}{x_i} \right) \\ 
  & \quad + \frac{1}{Re}\left(\pd{w_i}{x_j}, \,\tetak {\pd{\tvi^{c,k}}{x_j}} \right) -
\left(w_i,\frac{\trhok \, \widehat{g_i}}{Fr}\right) = 0, 
		\label{eqn:nav_stokes_var_semi_disc_vms}
	\end{split} \\
        \text{Thermo Consistency:} & \quad \tJi^{k} = \frac{\left(\rho_- - \rho_+ \right)}{2 \rho_{+}Cn} \, \pd{\tmu^{k}}{x_i}, \label{eqn:thermo_consistency_semi_disc_vms} \\
	\text{Chemical Potential:} & \quad 
	-\left(q,\tmu^{k}\right) + \left(q, \tpsi'^{k} \right) + Cn^2 \left(\pd{q}{x_i}, \, {\pd{\tphi^{k}}{x_i}}\right) = 0, \label{eqn:mu_eqn_var_semi_disc_vms}\\
	\text{Cahn-Hilliard Eqn:} & \quad \left(q, \frac{\phi^{k+1} - \phi^k}{\delta t} \right) -
	\left(\pd{q}{x_i}, \, \widetilde{u_i}^{k} \tphi^{k} \right) + \frac{1}{PeCn} \left(\pd{q}{x_i}, \, {\pd{\tmu^{k}}{x_i}} \right) = 0,
	\label{eqn:phi_eqn_var_semi_disc_vms}\\
    \begin{split}     
        \text{Pressure Poisson:} & \quad \left(\pd{q}{x_i}, \frac{1}{\trhok} \pd{p^{k+1}}{x_i} \right)=\frac{-2}{ \delta t} \left(q, \pd{v_i^{c,k+1}}{x_i}\right) +\left(\pd{q}{x_i}, \frac{1}{\trhok} \pd{p^{k}}{x_i} \right)\\
 & \boxed{- \frac{2}{ \delta t} \left(\pd{q}{x_i}, \frac{1}{\trhok} \tau_m R_m(\trhok,\Tilde{\eta_k}, v_i^{c,k+1}, u_i^k, \phi^k, p^k) \right)}
 \label{eqn:pressure_poisson_var_semi_disc_vms}  
    \end{split}\\
        \begin{split}      
        \text{Velocity update:} & \quad \left(w_i, \trhok u_i^{k+1} \right) + \frac{\delta t}{2}\left(w_i, \pd{p^{k+1}}{x_i} \right) = \left(w_i, \trhok v_i^{k+1} \right) + \frac{\delta t}{2}\left(w_i, \pd{p^{k}}{x_i} \right) \\
  & \boxed{- \left(w_i, \tau_m R_m(\trhok,\Tilde{\eta_k}, v_i^{c,k+1}, u_i^k, \phi^k, p^k) \right)}
  \label{eqn:velocity_update_var_semi_disc_vms}
    \end{split}
\end{align}
$\forall \vec{w} \in \mathscr{P}\vec{H_0^{1,r}}(\Omega), \textit{ and } q \in \mathscr{P}H^{1,r}(\Omega)$, given $\vec{u^k}, \vec{u^{k-1}} \in \vec{H_0^{1,r}}(\Omega), \textit{ and } p^k, \phi^{k},\mu^{k} \in H^{1,r}(\Omega)$.
	\label{def:variational_form_vms} 
\end{definition}

The stabilization term in the pressure-Poisson equation~\cref{eqn:pressure_poisson_var_semi_disc_vms} closely resembles the pressure-stabilizing Petrov–Galerkin formulation for the Navier–Stokes equations but is applied in a distinct equation compared to the solenoidality condition. Additionally, note that the Cahn-Hilliard equation~\cref{eqn:phi_eqn_var_semi_disc_vms} excludes a stabilization term for the convective term. This omission arises because introducing upwind dissipation as a stabilization measure disrupts the mass conservation property of the phase-field equation, causing $\int_{\Omega} \phi , \mathrm{d}x$ to vary over time instead of remaining constant. However, the dissipation error inherent in the standard cG discretization diminishes with finer mesh resolutions, ensuring that the numerical scheme remains convergent while preserving the mass conservation property of the Cahn-Hilliard equation~\cref{eqn:phi_eqn_var_semi_disc_vms}.

\subsection{Adaptive mesh refinement} \label{subsec:AMR}
Adaptive mesh refinement (AMR) is a widely used technique in numerical simulations to enhance accuracy by dynamically increasing resolution in specific regions of interest within the computational domain~\citep{Harmon2020, Zhang2020, Yong2021}. This approach is particularly advantageous in simulations with highly localized phenomena, such as steep gradients or interfaces, where uniform meshes would result in prohibitive computational costs. AMR is particularly indispensable for 3D simulations, where the exponential growth in computational and memory demands of uniform meshes often renders such problems infeasible. AMR achieves substantial efficiency gains by concentrating computational resources in regions of interest while maintaining high solution fidelity.

One of the most prominent methods for AMR is octree-based mesh refinement, which has been extensively applied to solve various engineering problems~\citep{Saurabh2021, Saurabh2023, Xu2021, Kim2002}. An octree is a hierarchical tree data structure that discretizes the spatial domain into a series of nested grids. At its core, an octree represents the domain as a coarse grid that can be recursively refined by subdividing individual cells into eight smaller octants. When a cell is subdivided, its corresponding node in the tree gains eight child nodes, called leaves, which collectively represent smaller cubic regions in space. This hierarchical structure allows efficient octree traversal to identify the node containing a given point in the domain~\citep{Sundar2008}.

Octree-based discretization is widely favored in scientific computing due to its hierarchical and adaptive representation of the computational domain~\citep{Becker2000}. Moreover, its structured yet flexible framework balances accuracy and computational efficiency, particularly in simulations with complex geometries or highly deforming interfaces. Researchers have also made significant advances in parallelizing computations on octree meshes by optimizing data structures and minimizing unnecessary inter-processor communication, ensuring scalability for large-scale problems~\citep{Saurabh2023}. These developments make octree-based AMR an essential tool for tackling computational challenges in modern simulations.

\subsubsection{Hanging nodes and 2:1 balancing} \label{subsec:hanging-nodes}
This paper utilizes the \texttt{DENDRO}-KT library, an octree-based meshing and finite element solver framework designed for parallel scalability and efficiency. The library incorporates robust parallel data structures, making it highly effective for tackling large-scale scientific computing problems across diverse domains~\citep{Saurabh2021, Xu2021}. To distribute the computational domain across multiple processors while ensuring load balance and minimizing inter-processor communication, we employ a space-filling curve (SFC) for partitioning. The SFC algorithm organizes the octants to preserve spatial locality, which is crucial for reducing communication overhead. Readers interested in a detailed explanation of this partitioning approach are referred to~\citet{Sundar2008}.

Imposing constraints on the relative sizes of neighboring octants is essential in many computational applications, as it ensures numerical stability and improves interpolation accuracy~\citep{Bern1999}. In the \texttt{DENDRO}-KT framework, we enforce a 2:1 balancing condition during mesh refinement. This condition ensures that adjacent octants differ by no more than one level of refinement, which is particularly important for maintaining a consistent and stable numerical solution across the mesh. Enforcing the 2:1 balancing condition mitigates abrupt resolution transitions, thereby reducing numerical artifacts and improving solver convergence.

\begin{figure} [h]
  \centering
  \begin{subfigure}[b]{0.35\textwidth}
    \includegraphics[width=\textwidth]{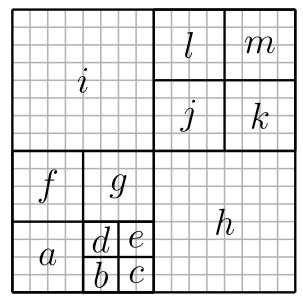}
    \caption{}
    \label{subfig:load_balancing_before}
  \end{subfigure}
  \begin{subfigure}[b]{0.35\textwidth}
    \includegraphics[width=\textwidth]{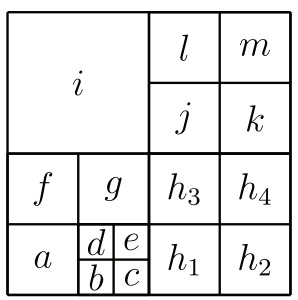}
    \caption{}
    \label{subfig:load_balancing_after}
  \end{subfigure}
  \caption{Illustration of 1:2 load balancing constraint in a 2D spatial domain \citep{Sundar2008}. (a) Before load balancing; (b) After load balancing (where the element $h$ is subdivided into $h_1,h_2,h_3,h_4$.)}
  \label{fig:octree_balancing}
\end{figure}

The 2:1 balancing limits hanging nodes to one per face or edge between adjacent octants. A hanging node is a degree of freedom introduced on the shared faces or edges of octants with different refinement levels~\citep{Saurabh2021}. Hanging node values are interpolated using basis functions of the coarser octant, ensuring solution continuity across varying resolution interfaces. This interpolation step is critical in maintaining the finite element solution's accuracy while accommodating the octree mesh's hierarchical structure. An example of 2:1 balancing is illustrated in 2D in \cref{fig:octree_balancing}, where the bottom right cell ($h$) is subdivided into four smaller cells to maintain the balancing condition. This hierarchical adjustment ensures the mesh refinement remains efficient and numerically stable while adapting to the problem's complexity.

\subsubsection{Top-down and bottom-up traversal}
An octree mesh is advantageous because it significantly reduces computational costs and memory requirements by adapting the mesh resolution to the problem's complexity. This flexibility dynamically allocates computational resources, ensuring fine resolution in critical regions (e.g., near interfaces or steep gradients) and coarser resolution elsewhere for optimal computational efficiency. Adaptive refinement is achieved by applying user-defined conditions to selectively subdivide specific regions in the space. Starting with a coarse grid featuring uniform octant sizes, the refinement process adopts a top-down approach where each octant is refined according to user-specified criteria. In our simulations, refinement is driven by the phase-field parameter $\phi$, ensuring high resolution near the interface where $\abs{\phi} \leq 0.9$, capturing the interfacial dynamics with greater accuracy.

Traversal-based assembly is a computationally efficient strategy for managing data in an octree without relying on traditional element-to-node maps. Our framework leverages the \texttt{DENDRO}-KT library, which optimizes the matrix and vector assembly through recursive top-down and bottom-up traversals of the tree structure. In top-down traversals, nodes are progressively propagated from coarser to finer levels of the tree, starting from the root. Information is inherited by finer octants, ensuring that all nodes within the adaptive mesh maintain consistency with their parent nodes. Nodes that overlap multiple subtrees are duplicated during this traversal to account for shared regions, maintaining solution continuity across the mesh. This recursive process continues until the traversal reaches the leaf nodes, where the finest resolution is applied. Conversely, bottom-up traversals aggregate values from finer to coarser levels, enabling efficient global matrix and vector assembly. Local contributions, such as stiffness matrices or solution vectors, are calculated at the leaf nodes and recursively passed upward to their parent nodes. This aggregation ensures that the global solution remains accurate and consistent across all octree levels. Boundary nodes and ghost elements facilitate communication between subtrees, enabling efficient data exchange for shared nodes. These mechanisms minimize communication overhead, improving scalability and parallel performance.

\begin{figure} 
  \centering
    \includegraphics[width=0.65\textwidth]{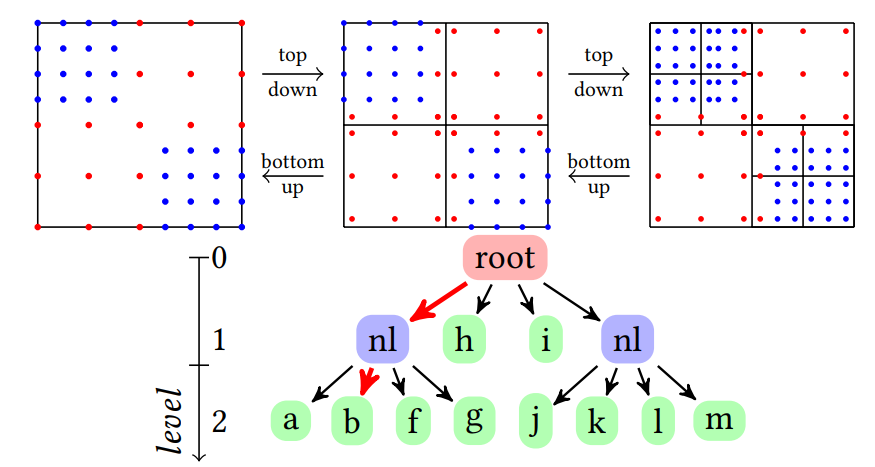}
  \caption{Illustration of top-down and bottom-up traversals in a 2D tree with quadratic element order \citep{Saurabh2021Scalable}. The leftmost figure shows the unique shared nodes (color-coded by tree level). During top-down traversal, shared nodes across children of a parent node are recursively duplicated, propagating downward until the leaf nodes are reached. At the leaf level, any missing local elemental nodes are interpolated from their immediate parent (as shown in the rightmost figure). The bottom-up traversal then aggregates the duplicated nodes, merging them into a consistent global structure.}
  \label{fig:octree_traversal}
\end{figure}

Combining top-down refinement and bottom-up aggregation ensures fine-grained resolution in critical regions while maintaining computational efficiency across the entire domain. This traversal strategy avoids the explicit storage of element-to-node maps, significantly reducing the memory footprint and enabling simulations with millions of degrees of freedom. Figure~\ref{fig:octree_traversal} illustrates the top-down and bottom-up processes, highlighting how information propagates dynamically through the octree hierarchy to achieve consistent and scalable assembly.

\subsubsection{\texttt{PETSc} solver}
We use the \texttt{PETSc} toolkit to solve linear and nonlinear systems using iterative solvers. \texttt{PETSc} stands for the Portable, Extensible Toolkit for Scientific Computation, and it is a versatile library widely used in scientific computing applications (see the manual in \citep{PETSc_manual}). Accessible from various object-oriented software platforms such as C++, it supports parallel linear and nonlinear solvers. As a highly scalable MPI-based library, \texttt{PETSc} is optimized for distributed memory architectures, enabling efficient parallelization and scalability for simulations involving millions of degrees of freedom~\citep{Balay2001}. This capability makes \texttt{PETSc} an excellent choice for our two-phase flow simulations, which involve computationally intensive fine-resolution interface regions and extreme density and viscosity ratios.

After conducting numerical experiments with various solvers, we found that the stabilized version of the Biconjugate Gradient (BiCGstab) solver, combined with Additive Schwarz preconditioning, is the most stable choice for our two-phase simulations. The full \texttt{PETSc} command line options are provided in \ref{petscCommands} for better reproducibility of our results. These options include solver configurations, preconditioner settings, and convergence criteria, allowing others to replicate our results accurately.
\definecolor{ForestGreen}{RGB}{34,139,34}
\section{Results}
\label{sec:results}

This section presents three canonical test cases, including the method of manufactured solutions, the capillary wave problem, and the single bubble rise in viscous flow. We then focus on the high-density ratio simulations of the oscillating droplet method. The velocity boundary conditions vary between problems and are specified in the corresponding section. We impose no-flux boundary conditions to the phase-field parameter $\phi$ and the chemical potential $\mu$, reflecting the assumption of an isolated system. The pressure is fixed to zero at a single reference point within the domain at $(0,0)$. In our simulations, we adopt a non-degenerate mobility model, where the nondimensional mobility $m(\phi)$ is treated as a constant, set to a value of one.

\subsection{Method of manufactured solutions} \label{subsec:manfactured_soln_result}
We employ the method of manufactured solutions to evaluate the order of accuracy of our numerical method. This involves selecting analytical solutions for our unknowns and calculating the corresponding residuals by substituting them into the Cahn-Hilliard Navier-Stokes equations. The resulting residuals effectively become forcing terms in our equations. Subsequently, we will solve for the unknowns and calculate their $L^2$ error compared to the analytical solutions. To achieve this, we chose the following solutions, 
\begin{equation}
	\begin{split}
		& \quad u = \sin(\pi x_1)\cos(\pi x_2)\sin(t) \\
		& \quad v = -\cos(\pi x_1)\sin(\pi x_2)\sin(t) \\
		& \quad p = \sin(\pi x_1)\sin(\pi x_2)\cos(t) \\ 
		& \quad \phi = \cos(\pi x_1)\cos(\pi x_2)\sin(t)\\
		& \quad \mu = \cos(\pi x_1)\cos(\pi x_2)\sin(t)
	\end{split}
	\label{eq:manufac_analytical}
\end{equation}

\begin{figure} %[b]
	\begin{subfigure}{0.48\textwidth}
		\centering
		\begin{tikzpicture}
			\begin{loglogaxis}[width=0.95\linewidth, scaled y ticks=true,xlabel={timestep},
				ylabel={$||u - u_{exact}||_{L^2}$},legend entries={$u$,$v$,$p$,$\phi$, $\mu$, $slope = 1$, $slope = 2$},
				legend style={nodes={scale=0.45, transform shape}}, legend pos=south east]
				\addplot table [x={timestep},y={L2U},col sep=comma] {Data/manufacturedSolution/time_lvl-8.csv};
				\addplot table [x={timestep},y={L2V},col sep=comma] 
				{Data/manufacturedSolution/time_lvl-8.csv};
				\addplot table [x={timestep},y={L2P},col sep=comma] 
				{Data/manufacturedSolution/time_lvl-8.csv};
				\addplot table [x={timestep},y={L2Phi},col sep=comma] 
				{Data/manufacturedSolution/time_lvl-8.csv};
				\addplot table [x={timestep},y={L2Mu},col sep=comma] 
				{Data/manufacturedSolution/time_lvl-8.csv};
				\addplot +[mark=none, red, dashed] [domain=0.01:1.0]{0.1*x};
				\addplot +[mark=none, blue, dashed] [domain=0.01:1.0]{0.025*x^2};
			\end{loglogaxis}
		\end{tikzpicture}
		\subcaption{}
		\label{subfig:manuf_sol_time_convergence}
	\end{subfigure}
	\begin{subfigure}{0.48\textwidth}
		\centering
		\begin{tikzpicture}
			\begin{loglogaxis}[width=0.95\linewidth, scaled y ticks=true,xlabel={Element Size $h$}, 
				ylabel={$||u - u_{exact}||_{L^2}$},legend entries={$u$,$v$,$p$,$\phi$, $\mu$, $slope = 1$, $slope = 2$},
				legend style={nodes={scale=0.45, transform shape}}, legend pos=south east]
				\addplot table [x={h},y={L2U},col sep=comma] {Data/manufacturedSolution/meshError_dt=5e-4.csv};
				\addplot table [x={h},y={L2V},col sep=comma] 
				{Data/manufacturedSolution/meshError_dt=5e-4.csv};
				\addplot table [x={h},y={L2P},col sep=comma] 
				{Data/manufacturedSolution/meshError_dt=5e-4.csv};
				\addplot table [x={h},y={L2Phi},col sep=comma] 
				{Data/manufacturedSolution/meshError_dt=5e-4.csv};
				\addplot table [x={h},y={L2Mu},col sep=comma] 
				{Data/manufacturedSolution/meshError_dt=5e-4.csv};;
				\addplot +[mark=none, red, dashed] [domain=0.003:0.4]{0.1*x};
				\addplot +[mark=none, blue, dashed] [domain=0.003:0.4]{0.025*x^2};
			\end{loglogaxis}
		\end{tikzpicture}
		\subcaption{}
		\label{subfig:manuf_sol_mesh_convergence}
	\end{subfigure}
	\caption{\textit{Manufactured Solution:} Panel (a) Temporal convergence using $h=1/2^8$; (b) spatial convergence using $\delta t = 5 \times 10^{-4}$.}
	\label{fig:manuf_sol_convergence}
\end{figure}

We compute numerical solutions with the following non-dimensional parameters: $Re = 1$, $We = 1$, $Cn = 1$, $Fr = 1$, and $Pe = 0.3$. The density and viscosity ratios are set to be $\rho_{-}/\rho_{+} = 10^{-2}$ and $\nu_{-}/\nu_{+} = 10^{-1}$. All $L^2$ errors are calculated at $t = \pi$ to allow for one complete cycle. 

We use a uniform mesh with $256 \times 256$ quadrilateral elements to test the temporal accuracy of our numerical framework. As seen in \cref{subfig:manuf_sol_time_convergence}, all unknowns are second-order accurate in time except for pressure $p$. This is expected since our numerical method only uses pressure at the previous timestep. We show mesh convergence in \cref{subfig:manuf_sol_mesh_convergence} by varying the mesh size $h$ while using a fixed timestep $\delta t = 5 \times 10^{-4}$. Since we are using linear basis functions, all five unknowns are expected to be second-order accurate in mesh size, as shown by our results. The pressure error tappers off at smaller mesh sizes since the $L^2$ error saturates due to temporal error from the fixed timestep $\delta t = 5 \times 10^{-4}$. 

\subsection{Capillary wave problem}

In this subsection, we utilize the benchmark two-phase capillary wave problem to assess the accuracy of the proposed method \cite{Dong2012JCP}. The problem is set up as follows: two immiscible, incompressible fluids occupy an infinite domain, with the lighter fluid in the upper half and the heavier fluid in the lower half. The interface separating the two fluids is initially perturbed from its horizontal equilibrium position by a small amplitude sinusoidal wave, and oscillations begin at time $t=0$. The objective is to analyze the motion of the interface over time. In \cite{Prosperetti1977}, an exact time-dependent standing wave solution for this problem was derived, assuming the fluids have matching kinematic viscosities while allowing their densities and dynamic viscosities to differ. The analytical expression for the decay of the wave amplitude, $\eta(t)$, is given as follows:

\begin{align}
    \frac{\eta(t)}{\eta_0} &= \frac{4(1-4\gamma)\nu^2 k^4}{8(1-4\gamma)\nu^2 k^4 + \omega_0} \, \text{erfc}(\sqrt{\nu k^2 t}) + \sum_{i=1}^{4} \frac{z_i}{Z_i} \frac{\omega_0^2}{z_i^2 - \nu k^2} e^{(z_i^2 - \nu k^2)t} \, \text{erfc}(z_i\sqrt{\nu t})
    \label{eq:decay_wave_amplitude}
\end{align}

Here, $\omega_0 = \sqrt{\frac{\sigma k^3}{\rho_H + \rho_L}}$ is the angular frequency, $\gamma = \frac{\rho_H \rho_L}{(\rho_H + \rho_L)^2}$, and $Z_i = \prod_{\substack{1 \le j \le 4 \\ j \ne i}} (z_j - z_i)$. The complementary error function $\text{erfc}(z_i)$ can be evaluated by solving the following algebraic equation:

\begin{equation}
    z^4 - 4\gamma \sqrt{\nu k^2} z^3 + 2 (1 - 6\gamma) \nu k^2 z^2 + 4 (1 - 3\gamma) (\nu k^2)^{3/2} z + (1 - 4\gamma) \nu k^2 + \omega_0^2 = 0.
    \label{eq:algebraic_equation}
\end{equation}

\begin{figure} %[ht]
    \centering
    \includegraphics[width=0.99\linewidth,trim=0 0 100 0,clip]{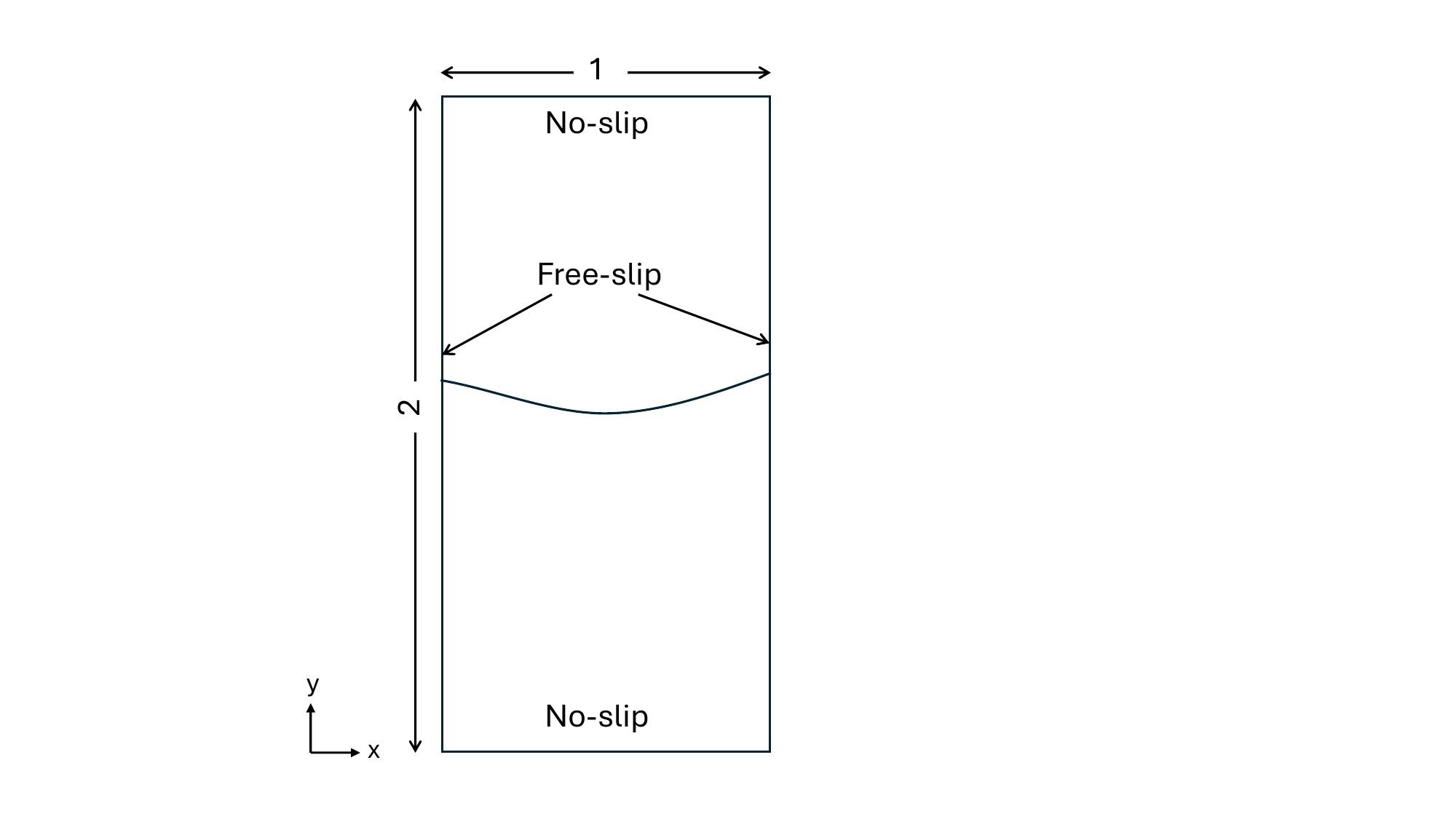}
    \caption{Schematic diagram of the capillary wave problem: computational domain and boundary conditions.}
    \label{fig:capillary_wave_setup}
\end{figure}

We simulate the problem with two phases of equal kinematic viscosity to validate our numerical solution and compare the results with the exact solution from \cite{Prosperetti1977}. The numerical setup is shown in Figure~\ref{fig:capillary_wave_setup}, with a computational domain of size $[0, 1] \times [0, 2]$. A no-slip boundary condition is applied to the top and bottom boundaries velocity, and free-slip conditions are enforced on the side walls, as illustrated in Figure~\ref{fig:capillary_wave_setup}. The equilibrium position of the interface aligns with the $x$-axis, and the initial perturbation of the interface is described by $y = H_0 \cos(2\pi x)$, where $H_0 = 0.01$ represents the initial amplitude. Given that the initial amplitude $H_0$ is small relative to the vertical dimension of the domain, the effect of the top and bottom walls on the interface motion is negligible.

We first test the case of an equal density ratio $\rho_+ / \rho_- = 1$ in zero gravity ($\text{Fr}=0$). The nondimensional numbers are chosen as: $\text{We}=1$, $\text{Cn}=0.005$, and $\text{Cn}=13,333.33$. Figure~\ref{fig:capillary_wave_1To1} compares numerical and analytical solutions at four Reynolds numbers ($\text{Re}=50$, $100$, $200$, and $500$), with equal kinematic viscosities, $\eta_+ / \rho_+ = \eta_- / \rho_-$. The time step used in the simulations is $\delta t = \num{2e-4}$. The mesh is uniform along the $x$ direction, while in the $y$ direction, it is clustered within the region $0.5 \leq y \leq 1.5$, with a mesh size ranging from $h = 4.0/2^8$ to $h = 4.0/2^{11}$, ensuring approximately 8 elements per interface. As shown in Figure~\ref{fig:capillary_wave_1To1}, the numerical simulations closely match the theoretical predictions for all Reynolds numbers tested. This agreement demonstrates the robustness of the proposed numerical framework in accurately capturing the physical behavior of capillary waves under varying flow conditions. Using a finer mesh in the interfacial region mitigates the effects of hanging nodes, thus providing a more stable solution (see \ref{subsec:hanging-nodes}). The results demonstrate the accuracy of the numerical framework, which aligns well with the analytical solution presented in \cite{Prosperetti1977}, validating its applicability to complex interfacial dynamics problems.

Next, we analyze the behavior of the proposed numerical framework under varying density ratios at a fixed Reynolds number, $\text{Re}=100$. Figure~\ref{fig:capillary_wave_Re=100} presents comparisons of the numerical and analytical solutions of the capillary wave problem using density ratios ($\rho_+ / \rho_-$) of $1$, $10$, $100$, and $10,000$. We use a time step of $\delta t = \num{2e-4}$ and a mesh size ranging from $h = 4.0/2^8$ to $h = 4.0/2^{11}$, as in the previous case. However, for the high density ratio case ($\rho_+ / \rho_- = 10,000$), the mesh is further refined to $h = 4.0/2^{12}$ to $h = 4.0/2^{11}$, ensuring 16 elements per interface. This increased resolution is essential for capturing steep gradients in material properties (density and viscosity), interfacial velocity, and pressure while reducing numerical instabilities associated with high-density ratio flows. As shown in Figure~\ref{fig:capillary_wave_Re=100}, the numerical results closely align with the analytical predictions in all density ratios, demonstrating the robustness and precision of the numerical method. The case of $\rho_+ / \rho_- = 10,000$ is particularly significant, as it underscores the ability of the method to handle extreme density contrasts, a primary focus of this study. These results highlight the framework's capability to accurately model interfacial dynamics, even under challenging physical conditions, aligning with the analytical solution.

\begin{figure}
    \centering
    % Row 1
    \begin{subfigure}{0.48\linewidth}
        \centering
        \begin{tikzpicture}[spy using outlines={rectangle, magnification=3, size=1cm, connect spies}]
            \begin{axis}[width=0.9\linewidth, xlabel={$t$ (s)}, ylabel={$H(t)$},
                xmin=0, xmax=5, ymin=-0.015, ymax=0.015, legend style={font=\tiny, at={(0.98,0.98)}, anchor=north east}]
                \addplot[red, only marks, mark size=0.5] table[x=time, y=amp, col sep=comma] {Data/capillaryWave/1To1/numerical_amp_Re=50.csv};
                \addplot[ForestGreen] table[x=time, y=amp, col sep=comma] {Data/capillaryWave/1To1/analytical_amp_Re=50.csv};
                \legend{Simulation, Theory}
    \coordinate (a) at (axis cs:1.25,0.0018); 
\end{axis}
\spy [black] on (a) in node at (2.5,3.5);         \end{tikzpicture}
        \subcaption{}
        \label{subfig:Re50}
    \end{subfigure}
    \begin{subfigure}{0.48\linewidth}
        \centering
        \begin{tikzpicture}[spy using outlines={rectangle, magnification=3, size=1cm, connect spies}]
            \begin{axis}[width=0.9\linewidth, xlabel={$t$ (s)}, ylabel={$H(t)$},
                xmin=0, xmax=5, ymin=-0.015, ymax=0.015, legend style={font=\tiny, at={(0.98,0.98)}, anchor=north east}]
                \addplot[red, only marks, mark size=0.5] table[x=time, y=amp, col sep=comma] {Data/capillaryWave/1To1/numerical_amp_Re=100.csv};
                \addplot[ForestGreen] table[x=time, y=amp, col sep=comma] {Data/capillaryWave/1To1/analytical_amp_Re=100.csv};
                \legend{Simulation, Theory}
    \coordinate (a) at (axis cs:1.2,0.003); 
\end{axis}
\spy [black] on (a) in node at (2.5,3.5);         \end{tikzpicture}
        \subcaption{}
        \label{subfig:Re100}
    \end{subfigure} \\
    % Row 2
    \begin{subfigure}{0.48\linewidth}
        \centering
        \begin{tikzpicture}[spy using outlines={rectangle, magnification=3, size=1cm, connect spies}]
            \begin{axis}[width=0.9\linewidth, xlabel={$t$ (s)}, ylabel={$H(t)$},
                xmin=0, xmax=5, ymin=-0.015, ymax=0.015, legend style={font=\tiny, at={(0.98,0.98)}, anchor=north east}]
                \addplot[red, only marks, mark size=0.5] table[x=time, y=amp, col sep=comma] {Data/capillaryWave/1To1/numerical_amp_Re=200.csv};
                \addplot[ForestGreen] table[x=time, y=amp, col sep=comma] {Data/capillaryWave/1To1/analytical_amp_Re=200.csv};
                \legend{Simulation, Theory}
    \coordinate (a) at (axis cs:1.2,0.0042); 
\end{axis}
\spy [black] on (a) in node at (2.5,3.7);         \end{tikzpicture}
        \subcaption{}
        \label{subfig:Re200}
    \end{subfigure}
    \begin{subfigure}{0.48\linewidth}
        \centering
        \begin{tikzpicture}[spy using outlines={rectangle, magnification=3, size=1cm, connect spies}]
            \begin{axis}[width=0.9\linewidth, xlabel={$t$ (s)}, ylabel={$H(t)$},
                xmin=0, xmax=5, ymin=-0.015, ymax=0.015, legend style={font=\tiny, at={(0.98,0.98)}, anchor=north east}]
                \addplot[red, only marks, mark size=0.5] table[x=time, y=amp, col sep=comma] {Data/capillaryWave/1To1/numerical_amp_Re=500.csv};
                \addplot[ForestGreen] table[x=time, y=amp, col sep=comma] {Data/capillaryWave/1To1/analytical_amp_Re=500.csv};
                \legend{Simulation, Theory}
    \coordinate (a) at (axis cs:1.2,0.0052); 
\end{axis}
\spy [black] on (a) in node at (2.5,3.8);         \end{tikzpicture}
        \subcaption{}
        \label{subfig:Re500}
    \end{subfigure}
    \caption{Comparison of the numerical and analytical capillary wave amplitudes for $\rho_+ / \rho_- = 1$ at different Reynolds numbers. Panel a) shows results for $Re=50$; panel b) shows results for $Re=100$; panel c) shows results for $Re=200$; and panel d) shows results for $Re=500$.}
    \label{fig:capillary_wave_1To1}
\end{figure}

\begin{figure}
    \centering
    % Row 1
    \begin{subfigure}{0.48\linewidth}
        \centering
        \begin{tikzpicture}[spy using outlines={rectangle, magnification=3, size=1cm, connect spies}]
            \begin{axis}[width=0.9\linewidth, xlabel={$t$ (s)}, ylabel={$H(t)$},
                xmin=0, xmax=5, ymin=-0.015, ymax=0.015, legend style={font=\tiny, at={(0.98,0.98)}, anchor=north east}]
                \addplot[red, only marks, mark size=0.5] table[x=time, y=amp, col sep=comma] {Data/capillaryWave/1To1/numerical_amp_Re=100.csv};
                \addplot[ForestGreen] table[x=time, y=amp, col sep=comma] {Data/capillaryWave/1To1/analytical_amp_Re=100.csv};
                \legend{Simulation, Theory}
    \coordinate (a) at (axis cs:1.2,0.003); 
\end{axis}
\spy [black] on (a) in node at (2.5,3.8);         \end{tikzpicture}
        \subcaption{}
        \label{subfig:1To1}
    \end{subfigure}
    \begin{subfigure}{0.48\linewidth}
        \centering
        \begin{tikzpicture}[spy using outlines={rectangle, magnification=3, size=1cm, connect spies}]
            \begin{axis}[width=0.9\linewidth, xlabel={$t$ (s)}, ylabel={$H(t)$},
                xmin=0, xmax=5, ymin=-0.015, ymax=0.015, legend style={font=\tiny, at={(0.98,0.98)}, anchor=north east}]
                \addplot[red, only marks, mark size=0.5] table[x=time, y=amp, col sep=comma] {Data/capillaryWave/highDensity_Re=100/numerical_amp_1To1e-1.csv};
                \addplot[ForestGreen] table[x=time, y=amp, col sep=comma] {Data/capillaryWave/highDensity_Re=100/analytical_amp_1To1e-1.csv};
                \legend{Simulation, Theory}
    \coordinate (a) at (axis cs:0.85,0.0045); 
\end{axis}
\spy [black] on (a) in node at (2.5,3.8);         \end{tikzpicture}
        \subcaption{}
        \label{subfig:1To1e-1}
    \end{subfigure} \\
    % Row 2
    \begin{subfigure}{0.48\linewidth}
        \centering
        \begin{tikzpicture}[spy using outlines={rectangle, magnification=3, size=1cm, connect spies}]
            \begin{axis}[width=0.9\linewidth, xlabel={$t$ (s)}, ylabel={$H(t)$},
                xmin=0, xmax=5, ymin=-0.015, ymax=0.015, legend style={font=\tiny, at={(0.98,0.98)}, anchor=north east}]
                \addplot[red, only marks, mark size=0.5] table[x=time, y=amp, col sep=comma] {Data/capillaryWave/highDensity_Re=100/numerical_amp_1To1e-2.csv};
                \addplot[ForestGreen] table[x=time, y=amp, col sep=comma] {Data/capillaryWave/highDensity_Re=100/analytical_amp_1To1e-2.csv};
                \legend{Simulation, Theory}
    \coordinate (a) at (axis cs:0.8,0.005); 
\end{axis}
\spy [black] on (a) in node at (2.5,3.8);         \end{tikzpicture}
        \subcaption{}
        \label{subfig:1To1e-2}
    \end{subfigure}
    \begin{subfigure}{0.48\linewidth}
        \centering
        \begin{tikzpicture}[spy using outlines={rectangle, magnification=3, size=1cm, connect spies}]
            \begin{axis}[width=0.9\linewidth, xlabel={$t$ (s)}, ylabel={$H(t)$},
                xmin=0, xmax=5, ymin=-0.015, ymax=0.015, legend style={font=\tiny, at={(0.98,0.98)}, anchor=north east}]
                \addplot[red, only marks, mark size=0.5] table[x=time, y=amp, col sep=comma] {Data/capillaryWave/highDensity_Re=100/numerical_amp_1To1e-4.csv};
                \addplot[ForestGreen] table[x=time, y=amp, col sep=comma] {Data/capillaryWave/highDensity_Re=100/analytical_amp_1To1e-4.csv};
                \legend{Simulation, Theory}
    \coordinate (a) at (axis cs:0.8,0.0047); 
\end{axis}
\spy [black] on (a) in node at (2.5,3.8);        \end{tikzpicture}
        \subcaption{}
        \label{subfig:1To1e-4}
    \end{subfigure}
    \caption{Comparison of the numerical and analytical capillary wave amplitudes for different density ratios using $Re=100$. Panel a) shows results for $\rho_+ / \rho_- = 1$; panel b) shows results for $\rho_+ / \rho_- = 10$; panel c) shows results for $\rho_+ / \rho_- = 100$; and panel d) shows results for $\rho_+ / \rho_- = 10,000$.}
    \label{fig:capillary_wave_Re=100}
\end{figure}

\subsection{Single bubble rise problem}

This section evaluates our framework on the benchmark single bubble rise problem. When a bubble rises through a viscous fluid, it is influenced by buoyancy, drag, and surface tension forces, which determine its shape and velocity \cite{Bhaga1981, Amaya2010, Hua2007}. We validate our numerical method against two cases reported in the literature \cite{Aland2012, Hysing2009}.

The bubble, representing a lighter fluid, has an initial diameter of 1 and is positioned at coordinates (1,1). The computational domain spans $[0, 2] \times [0, 4]$, with no-slip boundary conditions imposed on the top and bottom and free-slip conditions on the lateral boundaries. For case 1, the dimensionless parameters are set to: $\text{Re} = 35$, $\text{Fr} = 1$, $\text{We} = 10$, and $\text{Cn} = 0.01$. The density and viscosity ratios are $\rho_+ / \rho_- = 10$ and $\eta_+ / \eta_- = 10$, respectively. We present results for using $\delta t = 5 \times 10^{-4}$ and mesh size of $h = 4.0/2^7$ to $h = 4.0/2^{10}$, corresponding to approximately 8 elements per interface. As seen in Figure~\ref{subfig:bubbleRise_We=10}, the results for $\delta t = 5 \times 10^{-4}$ align closely with the literature~\cite{Hysing2009}. 

In case 2, we examine a scenario with reduced interfacial tension, resulting in greater interface deformation \cite{Aland2012}. The dimensionless parameters for this case are: $\text{Re} = 35$, $\text{Fr} = 1$, and $\text{We} = 125$, with density and viscosity ratios of $\rho_+ / \rho_- = 1000$ and $\eta_+ / \eta_- = 100$. We present results using $\delta t = 5 \times 10^{-4}$ for two interfacial thicknesses: $\text{Cn} = 0.01$ and $\text{Cn} = 0.0025$. Adaptive mesh refinement was applied, with mesh sizes ranging from $h = 4.0/2^7$ to $h = 4.0/2^{10}$ and $h = 4.0/2^7$ to $h = 4.0/2^{12}$, respectively, corresponding to 8 elements per interface. As illustrated in Figure~\ref{subfig:bubbleRise_We=125}, our results are consistent with the findings in \cite{Aland2012}, and the smaller interfacial thickness captures finer satellite drops in the bubble’s tail. These results demonstrate the accuracy of our numerical method in resolving bubble interfaces and capturing small-scale dynamics as the interfacial thickness decreases.

\begin{figure} 
	\begin{subfigure}{0.48\linewidth}
		\centering
		\begin{tikzpicture}
			\begin{axis} [width=\linewidth, height=\linewidth, xlabel={$x$},ylabel={$y$},legend columns=1,legend style={font=\tiny},xmin=0, xmax=2,ymin=0, ymax=4]
				% \addplot [color=red, only marks, mark size=1.5]  table[x=x,
				% y=y,
				% col sep=comma,
				% each nth point=10,
				% filter discard warning=false] {Data/bubbleRise/We=10/We=10_dt=2.5e-3.csv};       
				\addplot [color=ForestGreen, only marks, mark size=1.5]  table[x=x,
				y=y,
				col sep=comma,
				each nth point=10,
				filter discard warning=false] {Data/bubbleRise/We=10/We=10_dt=5e-4.csv};  
    			\addplot [color=blue, only marks, mark size=1.5]  table[x=x,
				y=y,
				col sep=comma,
				each nth point=1,
				filter discard warning=false] {Data/bubbleRise/We=10/Hysing_We10.csv};  
				% \legend{$\delta t = 2.5 \times 10^{-3}$, $\delta t = 5 \times 10^{-4}$, Hysing et.al (2009)}
    			\legend{$Cn = 1 \times 10^{-2}$, Hysing et.al (2009)}
			\end{axis}
		\end{tikzpicture}
		\subcaption{}
		\label{subfig:bubbleRise_We=10}
	\end{subfigure} 
	\begin{subfigure}{0.48\linewidth}
		\centering
		\begin{tikzpicture}
			\begin{axis} [width=\linewidth, height=\linewidth, xlabel={$x$},ylabel={$y$},legend columns=1,legend style={font=\tiny},xmin=0, xmax=2, ymin=0, ymax=4]
				\addplot [color=red, only marks, mark size=1.5]  table[x=x,
				y=y,
				col sep=comma,
				each nth point=1,
				filter discard warning=false] {Data/bubbleRise/We=125/We=125_Cn=1e-2.csv};
				\addplot [color=ForestGreen, only marks, mark size=1.5]  table[x = x,
				y=y,
				col sep=comma,
				each nth point=10,
				filter discard warning=false] {Data/bubbleRise/We=125/We=125_Cn=2.5e-3.csv};
    			\addplot [color=blue, only marks, mark size=1.5]  table[x = x,
				y=y,
				col sep=comma,
				each nth point=1,
				filter discard warning=false] {Data/bubbleRise/We=125/Aland_Voight_We125.csv};
				\legend{$Cn = 1 \times 10^{-2}$, $Cn = 2.5 \times 10^{-3}$, Aland and Voigt (2012)}
			\end{axis}
		\end{tikzpicture}
		\subcaption{}
		\label{subfig:bubbleRise_We=125}
	\end{subfigure} 
	\caption{Comparison of the bubble shape at the nondimensional time $t=4.2$ between current simulations and the literature. Panel a) shows contours for case 1 using $Cn = 1 \times 10^{-2}$ compared to results in \citet{Hysing2009}; panel b) shows contours for case 2 using two different interfacial thicknesses ($Cn = 1 \times 10^{-2}$ and $Cn = 2.5 \times 10^{-3}$) compared to results in \citet{Aland2012}.}
 \label{fig:bubbleRise}
\end{figure}

\subsection{Problem setup}
\label{subsec:geom_bc_ic}
We compare the results from our simulations with the measured interfacial tension using the Oscillating Droplet Method (ODM). The principal problem in such numerical simulations is the large density ratio between the two phases~\citep{Zheng2022}. In the ODM process, the metal is melted by a laser before being levitated using an electrostatic levitator furnace (ELF), where the droplet weight is balanced by the electric force applied by the ELF, keeping the droplet at equilibrium. The levitated droplet is then excited using an external electric field superposed on top of the ELF electric field to transform from a spherical shape into an ellipsoid.  After shutting the external field, the excited droplet minimizes its surface energy by undergoing periodic damped oscillations to transition into its minimum energy spherical configuration.  The frequency of the oscillating drop $\omega$ can be related to the surface tension $\sigma$ through the following Rayleigh formula involving the metal density $\rho$ and radius $R_{o}$ of the sample,
 \begin{equation}
     \omega =\sqrt{\frac{(l)(l-1)(l+2)\sigma}{\rho {R_o}^3}},
     \label{eqn:rayleigh_general}
 \end{equation} 
\noindent where $l$ is the mode of oscillation. For mode-2 oscillations (oblate ellipsoids), Rayleigh's formula in \cref{eqn:rayleigh_general} simplifies to, 
 \begin{equation}
     {\omega_{o}}^2 =\frac{8\sigma_{o}}{\rho {R_o}^3}.
     \label{eqn:rayleigh_ellipse_3D}
 \end{equation}
%Due to the complexity and high cost of such microgravity experiments, developing a sound numerical framework that can accurately predict the oscillation behavior in ODM for realistic surface tensions and density contrasts can be very useful. Even though most of the available analysis in literature is focused on droplet oscillations in 3D~\citep{Sumaria2019} as it can be compared with experiments~\citep{Brenn1993}, an interfacial tension equation for 2D droplet oscillations has been developed in~\citet{Aalilija2020}. 2D oscillation models are cheaper and can be a good start in testing solver accuracy and stability before proceeding into the more expensive 3D simulations. The 2D droplet oscillations correspond to a transverse section of a droplet with no longitudinal variations. Assuming the oscillating droplet has a constant volume and using the energy balance equation, we can calculate the interfacial tension equation in terms of 2D oscillation frequency as follows~\citep{Aalilija2020}.

Given the complexity and high cost of conducting microgravity experiments, establishing a robust numerical framework to accurately predict oscillatory behavior in the oscillating droplet method (ODM) under realistic surface tensions and density contrasts is highly beneficial. While much of the existing literature focuses on droplet oscillations in 3D \citep{Sumaria2019}, largely due to their direct comparison with experimental data \citep{Brenn1993}, a 2D interfacial tension model for droplet oscillations was introduced by \citet{Aalilija2020}. Two-dimensional oscillation models offer a cost-effective alternative and provide a practical foundation for testing solver accuracy and stability before undertaking more computationally intensive 3D simulations. In the 2D case, the oscillation represents a transverse cross-section of a droplet with no longitudinal variation. Using the mass and energy conservation of the oscillating droplet, the interfacial tension equation can be derived in terms of the 2D oscillation frequency, as shown below \citep{Aalilija2020}:

 \begin{equation}
     {\omega_{o}}^2 =\frac{6\sigma_{o}}{\rho {R_o}^3}.
     \label{eqn:rayleigh_ellipse_2D}
 \end{equation}
 
For all the oscillating droplet method simulations in the paper, we consider a cubic domain and a metal droplet initially placed at the center of the domain. The droplet is perturbed from its equilibrium spherical shape into an ellipsoid to represent the oscillating droplet method. To represent this shape, we initialize the phase field variable $\phi$ using a $\tanh(\vec{x})$ function, which models the transition across the interface between the inside and outside of the droplet. The $\tanh$ function defines the diffusion layer across the droplet's boundary, while the underlying shape of the droplet is prescribed as an ellipse (or ellipsoid in 3D). This configuration represents the droplet shape following the initial excitation from the external field. 

The major diameter of the droplet, $D$, serves as the characteristic length scale in our problem and is assumed to be $2.5$ \si{\milli\meter}, consistent with typical microgravity experiments. To determine the characteristic velocity scale, we fix the Reynolds number at 300, which effectively defines a velocity scale $u_r$ that can be used to nondimensionalize and scale our results back to the physical, dimensional space. %By fixing the Reynolds number, we ensure that the flow regime remains consistent, and extensive numerical simulations confirm that this choice does not compromise the generality of our findings. 
The Reynolds and Weber numbers of the different systems studied in this paper are reported in~\cref{tab:nondimenNumbers}. Specifically, we consider molten droplets of Osmium, Copper, and Zirconium, which are canonical materials that have been tested using the ODM and exhibit very high densities with respect to the surrounding medium (air/helium).

\subsection{Setup of ODM numerical experiments}
\label{subsec:sim_setup}

We study three systems with varying density ratios, \textit{viz.} Zirconium-Helium, Copper-Air, and Osmium-Helium. We select the thermo-physical properties of the metal at their melting points. \Cref{tab:materialProperties} shows the material properties while \cref{tab:nondimenNumbers} shows the non-dimensional numbers and the density (viscosity) ratios used in the simulations. Note that these simulations are done in zero gravity, and thus the $\left(w_i,\frac{\trhok \, \widehat{g_i}}{Fr}\right)$ in the velocity prediction equation is omitted. A no-slip boundary condition is imposed on the mixture velocity $\vec{v}$. 

%We found that tuning the stabilization parameter \eqref{eqn:tau_m} with $s = 10^{-2}$ provided the most stable results for our simulations. This adjustment maintains numerical stability, providing a balanced approach for high-density ratio simulations.

We postprocess the simulation results to compute the oscillation amplitude $A(t)$ to understand the damped oscillation dynamics. $A(t)$ is defined as the distance between the two nodes on the interface corresponding to $\phi = 0$. This can be calculated by monitoring the phase field $\phi$ along a vertical line through the droplet center at every time step.  $A(t)$ is calculated by linear interpolation of $\phi$ versus $y$ values along the vertical line.

To appropriately resolve the interface dynamics, a small enough interface thickness $\epsilon$ must be chosen by specifying the Cahn number $Cn=\frac{\epsilon}{L_r}$. This choice can be accomplished by running numerical experiments for different interface thicknesses where $A(t)$ converges as we keep decreasing $Cn$. We select the largest $Cn$ for which the change in $A(t)$ is negligible for subsequently smaller $Cn$. Using this process, we chose the $Cn = \num{1e-2}$. It is important to note that decreasing $Cn$ further would result in a more expensive mesh \footnote{Accurately resolving the interface requires a sufficient number of elements, so thinner interfaces need finer meshes.}, and a smaller timestep. 
%We choose the major diameter of the droplet $D$ (assumed to be $2.5$ \si{\milli\meter}) as the non-dimensionalizing length scale. The schematic of the computational domain and initial condition of the metal droplet is shown in~\cref{fig:mesh}. In our case, choosing a non-dimensional velocity and time-scale is not obvious. For this reason, we choose the non-dimensional time-scale such that $Re = 300$. For that fixed $Re$, we calculate the Weber number ($We$) accordingly. 
%Note that the choice of scaling Reynolds number does not affect our results.

\begin{table}[H] \centering
\begin{tabular}{@{}|c|c|c|c|@{}}
\toprule
{Metal}	    &  {Interfacial Tension $\sigma$ (\si{\newton\per\meter})} & {Specific density (\si{\kilogram\per\meter\cubed})} & {Viscosity (\si{\pascal\second})} \\ 
\midrule
\midrule
{Osmium}    & {$2.7$}                                              & {$19100$}                                           & {\num{7e-3}}                      \\
{Zirconium} & {$1.45$}                                            & {$5892$}                                         & {\num{4.19e-3}}                   \\
{Copper}    & {$1.16$}                                             & {$7507$}                                            & {\num{2.01e-3}}                   \\
\bottomrule
\end{tabular}
\caption{Physical properties of the selected metals at their melting points \cite{Ishikawa2017, Marcus2016, Sumaria2019, Sumaria2017}.}
\label{tab:materialProperties} 
\end{table}

\begin{table}[H] \centering
\begin{tabular}{@{}|c|c|c|c|c|@{}}
\toprule
{System}           & {$We$}   & {$Re$}     & {${\rho_{+}}/{\rho_{-}}$} &{ ${\eta_{+}}/{\eta_{-}}$}  \\ 
\midrule
\midrule
{Osmium-Helium}    & {\num{3.42e-2}}    & {$300$} & {\num{117.54e3}}          & {$350.0$} 					\\
{Zirconium-Helium} & {\num{7.39e-2}}    & {$300$}  & {\num{36.26e3}}           & {$209.7$}  					\\
{Copper-Air}       & {\num{1.67e-2}}    & {$300$} & {\num{6.34e3}}            & {$111.0$}      				\\
\bottomrule
\end{tabular}
	\caption{Non-dimensional numbers, along with density and viscosity ratios, for the systems under consideration ($Fr=0$).}
	\label{tab:nondimenNumbers}                            
\end{table}

\subsection{Mesh Configuration}\label{subsec:mesh_configuration}

\begin{figure} [h]
	\centering
    \includegraphics[width=0.8\textwidth]{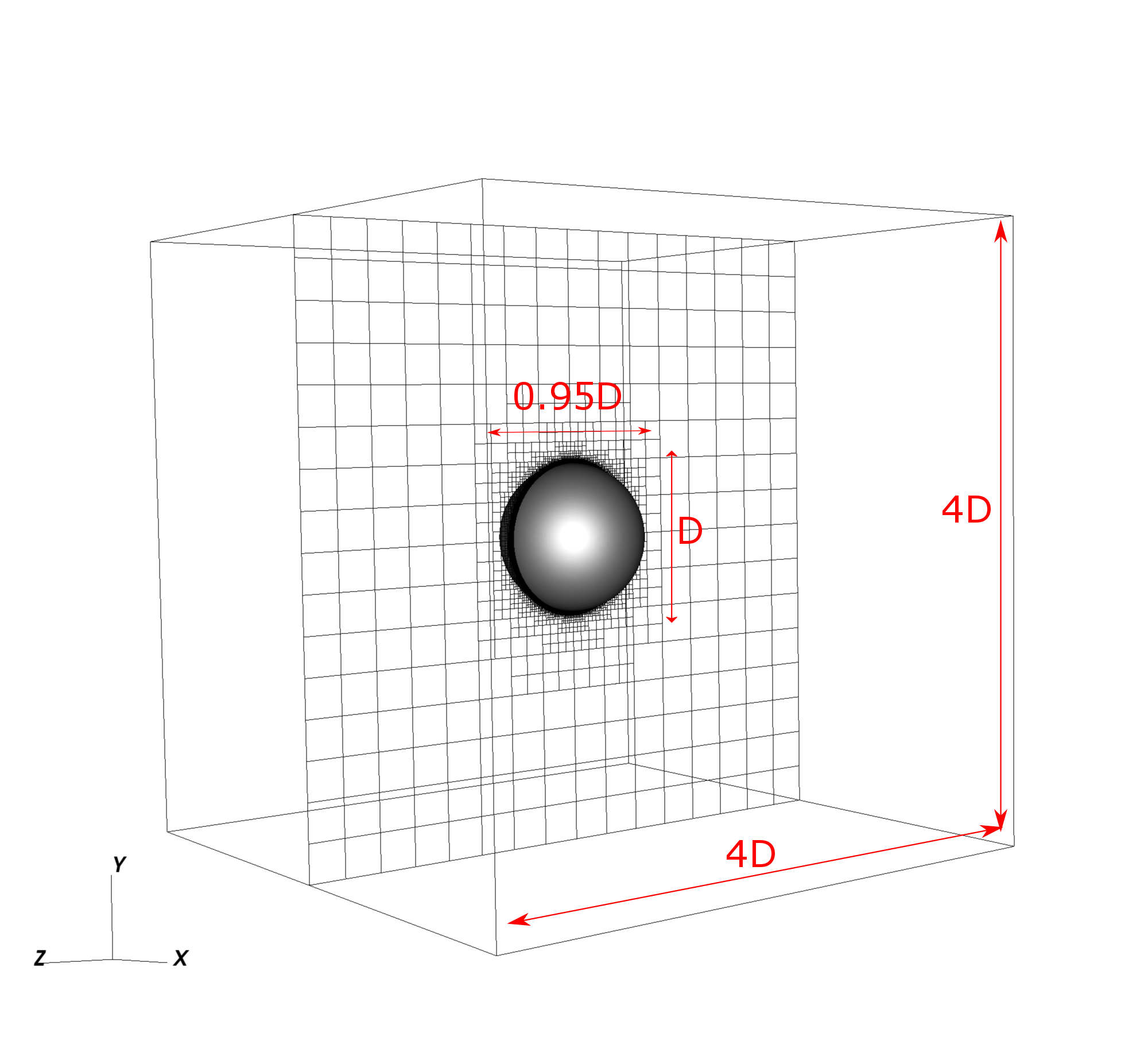}
   	\caption{Initial configuration of the levitated droplet inside the computational domain with a 2D slice of the mesh used in our simulations.}
   	\label{fig:mesh}
\end{figure}

Below is a schematic drawing of the levitated droplet in the gas environment and a slice of the mesh used for some of our test cases. The shown mesh in~\cref{fig:mesh} includes three refinement levels $l_r$ (10, 8, and 4) corresponding to the interface, droplet oscillation boundary region, and background mesh, respectively. The outer bounding box is chosen large enough to make the velocity near the boundaries small, ensuring our theoretical energy decay results hold. For each refinement level, mesh size is defined as $h=4.0/2^{l_r}$. For this specific mesh, $h=4.0/2^{10}$ corresponds to 16 elements along the interface. We gradually coarsen the mesh away from the droplet interface, with the bounding square box around the droplet's initial position having a mesh resolution of $h=4.0/2^8$. Similarly, the background mesh is coarser with a larger mesh size of $h=4.0/2^4$. Using a fine mesh near the interface is important where material properties, namely density and viscosity, change quickly across the interface. Also, the mesh inside the droplet bounding box should be fine enough to resolve the dynamics around the droplet, such as large votricities and fast fluid movement near the interface (high velocity). The background mesh can be relatively coarse as the surrounding fluid has a much smaller density and viscosity than the droplet, and less dynamics happen away from the interface and closer to the domain boundaries. As seen in~\cref{fig:mesh}, the droplet's major diameter is a quarter of the box domain. Initially, the droplet is at 5\% ellipticity, corresponding to its excited state, before oscillating to its resting (spherical) shape. Note that the total number of nodes in this 3D mesh is 3.7 million nodes.
Note that only two refinement levels are needed in most of our simulations, but a third refinement level is sometimes necessary for the linear solver convergence and to resolve some small-scale dynamics in certain regions in the mesh. 

\subsection{Diffuse Interface Thickness}
For diffuse interface models like the model we are using in this paper, the solution tends to match the real
physics solution in the limit of Cn tending to zero. However, decreasing $Cn$ will decrease the interfacial thickness, making the problem harder to solve. To ensure compatibility of results, we used the same interfacial resolution for different $Cn$, meaning that we maintained 16 mesh points inside the interface for different interface thicknesses. We use the oscillation dynamics $A(t)$ as an indicator for comparing numerical solutions. The three curves in~\cref{fig:CnConvergence} correspond to the Zirconium-Helium system. All 3 cases are in-phase and characterized by equal periods but different damping. The difference in oscillation amplitude $A(t)$ decreases with the Cahn number showing convergence towards the analytical, sharp-interface solution with $Cn=\num{5e-3}$. It is important to note that decreasing the Cahn number makes the problem stiffer and harder to solve. In addition to the denser mesh, decreasing the Cahn number will require a smaller time-step to resolve the system dynamics correctly. In our Zirconium-Helium simulations, our solver converges using a relatively large time-step $\delta t=\num{1e-5}$ when using $Cn=\num{1e-2}$. However, for the case of $Cn=\num{5e-3}$, a smaller time-step $\delta t=\num{5e-6}$ is required ($\delta t=\num{2.5e-6}$ for the case of $Cn=\num{2.5e-3}$). Since we are focused on calculating the interfacial tension of our droplets, we will use $Cn=\num{1e-2}$ for further analysis in this paper as it produces the most computationally efficient solutions while still giving accurate oscillation frequency values.

\begin{figure} 
\centering 
\begin{tikzpicture}[spy using outlines={rectangle, magnification=2, size=1.5cm, connect spies}]
\begin{axis} [width=0.85\linewidth,xlabel={$t(s)$},ylabel={$A(t)$},legend columns=1,legend style={font=\tiny},xmin=0, xmax=0.025,ymin=0.94, ymax=1.015]
\addplot [color=red]  table[x=time_Cn_1e-2,
         y=amp_Cn_1e-2,
         col sep=comma,
         each nth point=5,
         filter discard warning=false] {Data/Cn,We=0.0835,16ElementsPerInterface,physicalTime.csv};
         
\addplot [color=ForestGreen]  table[x=time_Cn_5e-3,
         y=amp_Cn_5e-3,
         col sep=comma,
         each nth point=5,
         filter discard warning=false] {Data/Cn,We=0.0835,16ElementsPerInterface,physicalTime.csv};
         
\addplot [color=blue] table[x=time_Cn_2.5e-3,
         y=amp_Cn_2.5e-3,
         col sep=comma,
         each nth point=5,
         filter discard warning=false] {Data/Cn,We=0.0835,16ElementsPerInterface,physicalTime.csv};   
    \legend{$Cn=10^{-2}$, $Cn=5$x$10^{-3}$, $Cn=2.5$x$10^{-3}$}
    \coordinate (a) at (axis cs:0.007,0.9975); 
\end{axis}
\spy [black] on (a) in node at (1.7,6.5); 
\end{tikzpicture} 
    \caption{$Cn$ convergence: the impact of interfacial thickness on droplet shape oscillations. The background mesh size for all three curves is $h=4.0/2^{4}$. The interfacial mesh sizes are $h=4.0/2^{10}$, $h=4.0/2^{11}$, and $h=4.0/2^{12}$ for $Cn=10^{-2}$, $Cn=5\times10^{-3}$, and $Cn=2.5\times10^{-3}$, respectively, corresponding to 16 elements per interface.}
	\label{fig:CnConvergence}
\end{figure}

\subsection{Mesh and Time Convergence in 2D}

To understand the effect of mesh resolution and to choose the appropriate mesh for further analysis, we performed simulations at a small time-step of $\delta t=\num{1e-5}$. We choose the Zirconium-Helium system for our mesh convergence analysis. We vary the mesh resolution by specifying two resolution levels. The first level indicates the mesh size in the interfacial region, while the second level corresponds to the background mesh as discussed in~\cref{subsec:mesh_configuration}. We can see from the zoomed inset in~\cref{subfig:mesh_conv} that $A(t)$ does not change with further increasing the refinement level beyond the case of $h=4.0/2^4$ to $h=4.0/2^{10}$. Similarly, we performed simulations at the finest spatial resolution $h=4.0/2^8$ to $h=4.0/2^{11}$ to study the effect of time-step on the droplet oscillations. \Cref{subfig:time_conv} shows $A(t)$ as a function of time t in seconds for varying time steps. We can see from the zoomed inset of~\cref{subfig:time_conv} that $A(t)$ are very close for both time steps, and it is impossible to distinguish them visually. Therefore, decreasing the time-step beyond $\delta t=\num{2e-5}$ does not affect the accuracy of our droplet shape oscillations $A(t)$. 

\begin{figure} %[H]
	\begin{subfigure}{\linewidth}
		\centering
		\begin{tikzpicture}[spy using outlines={rectangle, magnification=4, size=0.75cm, connect spies}]
			\begin{axis} [width=0.55\linewidth, xlabel={$t(s)$},ylabel={$A(t)$},legend columns=1,legend style={font=\tiny},xmin=0, xmax=0.0221,ymin=0.94, ymax=1.015]
				\addplot [color=red]  table[x =time_11_8,
				y=mesh_11_8,
				col sep=comma,
				each nth point=10,
				filter discard warning=false] {Data/meshConvergence,We=0.0835,dt=1e-5,physicalTime.csv};         
				\addplot [color=ForestGreen]  table[x=time_10_8,
				y=mesh_10_8,
				col sep=comma,
				each nth point=10,
				filter discard warning=false] {Data/meshConvergence,We=0.0835,dt=1e-5,physicalTime.csv};
				\addplot [color=blue] table[x=time_10_4,
				y=mesh_10_4,
				col sep=comma,
				each nth point=10,
				filter discard warning=false] {Data/meshConvergence,We=0.0835,dt=1e-5,physicalTime.csv};           
				\legend{$h=4.0/2^{11}$ to $h=4.0/2^8$,$h=4.0/2^{10}$ to $h=4.0/2^8$,$h=4.0/2^{10}$ to $h=4.0/2^4$}
				\coordinate (a) at (axis cs:0.0107,0.951);
			\end{axis}
			\spy [black] on (a) in node at (2.5,0.95);
		\end{tikzpicture}
		\subcaption{}
		\label{subfig:mesh_conv}
	\end{subfigure} \\
	\begin{subfigure}{\linewidth}
		\centering
		\begin{tikzpicture}[spy using outlines={rectangle, magnification=4, size=0.75cm, connect spies}]
			\begin{axis} [width=0.55\linewidth, xlabel={$t(s)$},ylabel={$A(t)$},legend columns=1,legend style={font=\tiny},xmin=0, xmax=0.0221,ymin=0.94, ymax=1.015]
				\addplot [color=red]  table[x=time_dt_1e-5,
				y=amp_dt_1e-5,
				col sep=comma,
				each nth point=10,
				filter discard warning=false] {Data/timeConvergence,lvl-10-8,physicalTime.csv};
				\addplot [color=ForestGreen]  table[x = time_dt_2e-5,
				y=amp_dt_2e-5,
				col sep=comma,
				each nth point=10,
				filter discard warning=false] {Data/timeConvergence,lvl-10-8,physicalTime.csv};
				\legend{$dt = 10^{-5}$,$dt = 2$x$10^{-5}$}
				\coordinate (a) at (axis cs:0.0107,0.951);
			\end{axis}
			\spy [black] on (a) in node at (2.5,0.95);
		\end{tikzpicture}
		\subcaption{}
		\label{subfig:time_conv}
	\end{subfigure}
	\caption{Panel a) shows mesh convergence in 2D using $\delta t=\num{1e-5}$; panel b) shows time convergence in 2D using $h=4.0/2^{11}$ to $h=4.0/2^8$. }
\end{figure}

\subsection{Streamlines}
\begin{figure} 
	\centering
		%%%%%%%%%%%%%%%%%%%%%% 4 %%%%%%%%%%%%%%%%%%%%%%%%%%%%%%%%%%%
		\begin{subfigure}{0.45\textwidth}
				\centering
				\begin{tikzpicture}
				\begin{axis} [xlabel={$t(s)$},ylabel={$A(t)$},legend columns=2,xmin=0, xmax=0.0092,yticklabel style={/pgf/number format/precision=3}, legend style={at={(0.5,-0.2)},anchor=north}]
				\addplot [color=red]  table[x=time_ZrHe,
				y=amp_ZrHe,
                col sep=comma,
                each nth point=10,
                filter discard warning=false] {Data/systems_3D_Re=300.csv};
                \node[label={180:{$t_1$}},circle,fill,inner sep=2pt] at (axis cs:0.00135,0.973) {};
                \node[label={180:{$t_2$}},circle,fill,inner sep=2pt] at (axis cs:0.0043,0.957) {};
                \node[label={180:{$t_3$}},circle,fill,inner sep=2pt] at (axis cs:0.0061,0.997) {};
                \node[label={180:{$t_4$}},circle,fill,inner sep=2pt] at (axis cs:0.0084,0.946) {};
				\end{axis}
				\end{tikzpicture}
				\subcaption{}
				\label{subfig:zr_he_timeshots}
			\end{subfigure}\\
		%%%%%%
			\centering
			\begin{subfigure}{0.45\textwidth}	
			\centering
			\includegraphics[width=1.0\linewidth]{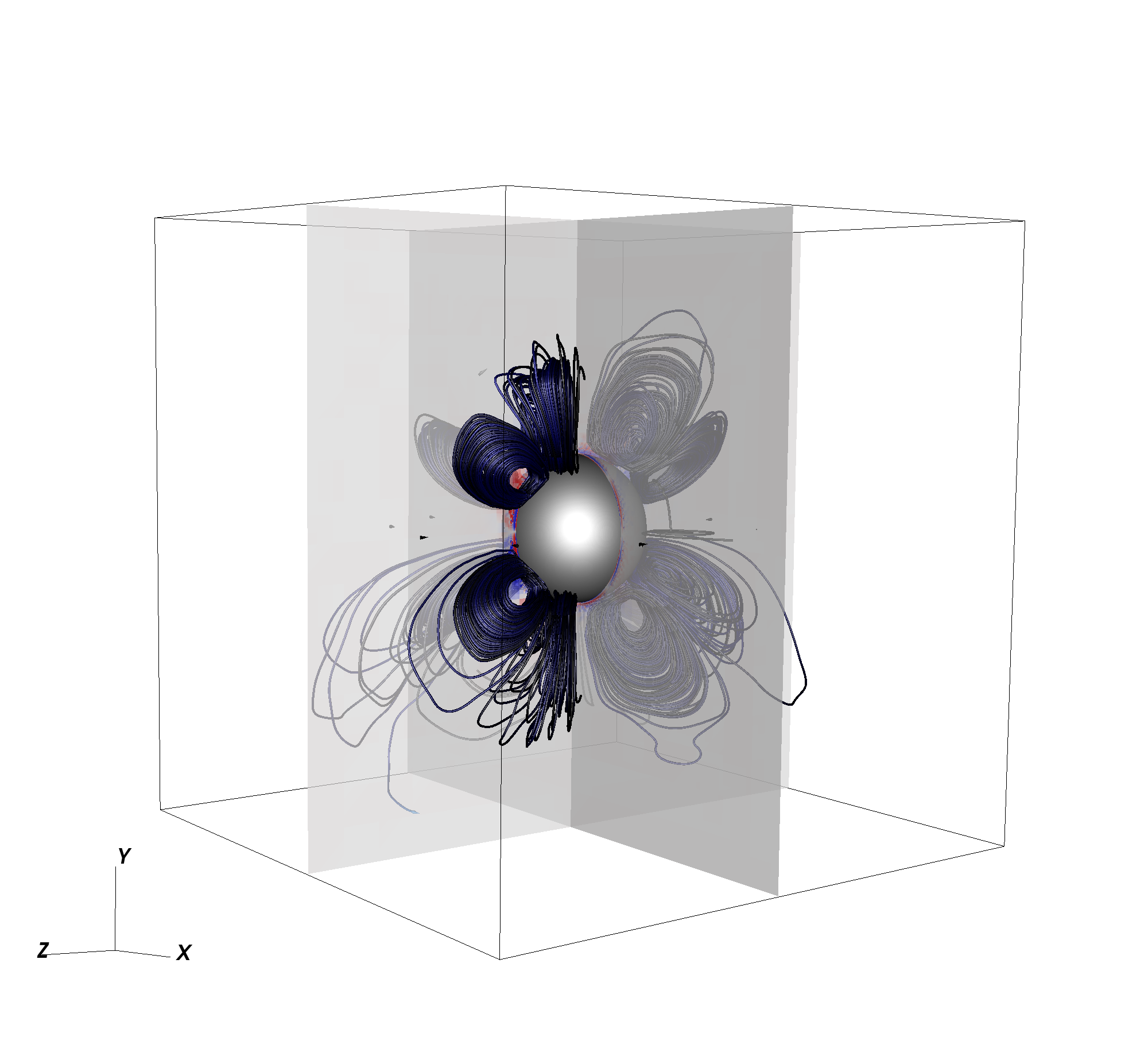}
			\subcaption{}
			\label{subfig:zr_he_t1}
			\end{subfigure}%
		%%%%%%%%%%%%%%%%%%%%%% 5 %%%%%%%%%%%%%%%%%%%%%%%%%%%%%%%%%%%
			\centering
			\begin{subfigure}{0.45\textwidth}	
			\centering
			\includegraphics[width=1.0\linewidth]{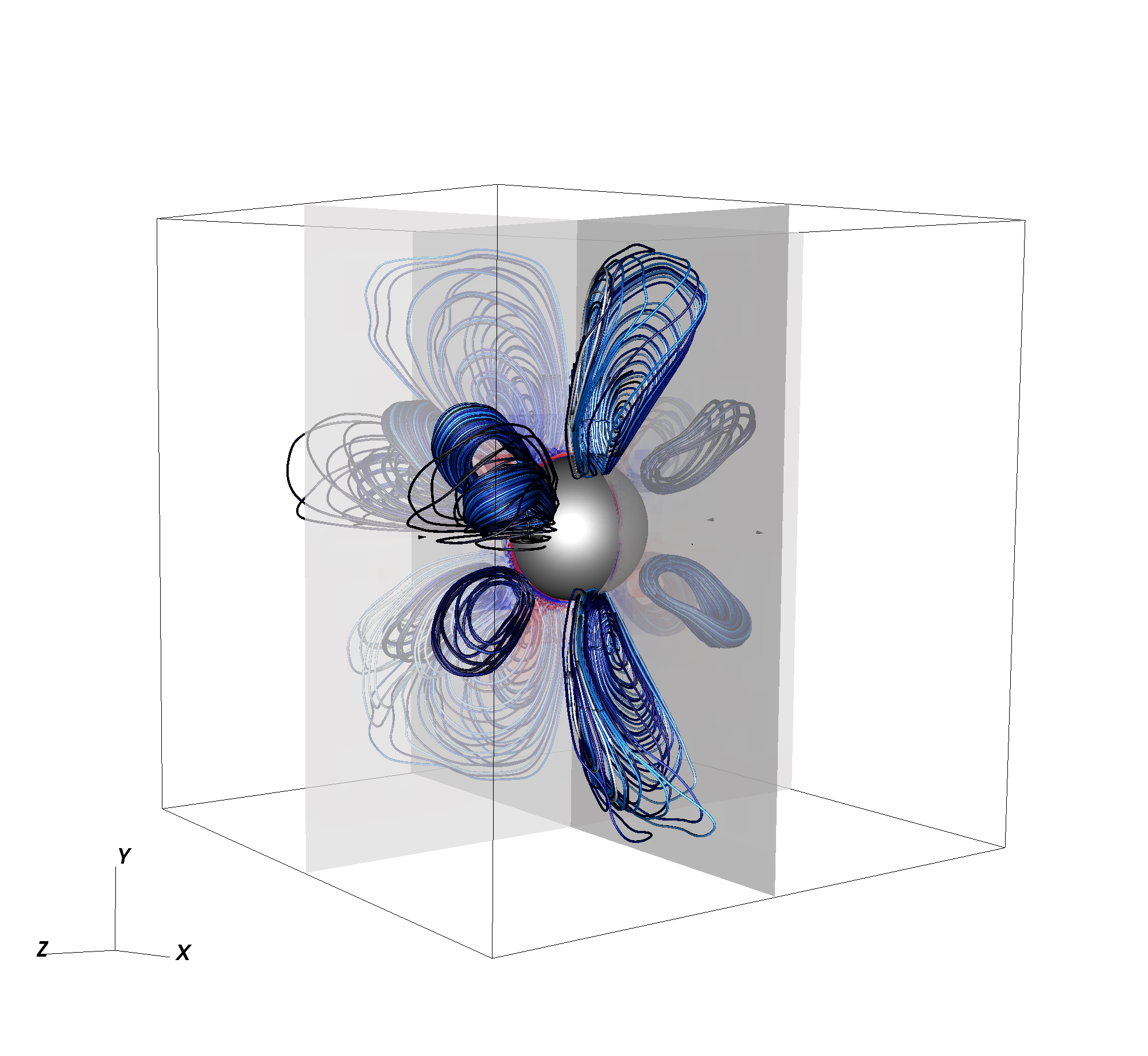}
                \subcaption{}
			\label{subfig:zr_he_t2}
			\end{subfigure}\\
		%%%%%%%%%%%%%%%%%%%%%% 6 %%%%%%%%%%%%%%%%%%%%%%%%%%%%%%%%%%%
			\centering
			\begin{subfigure}{0.45\textwidth}	
			\centering
			\includegraphics[width=1.0\linewidth]{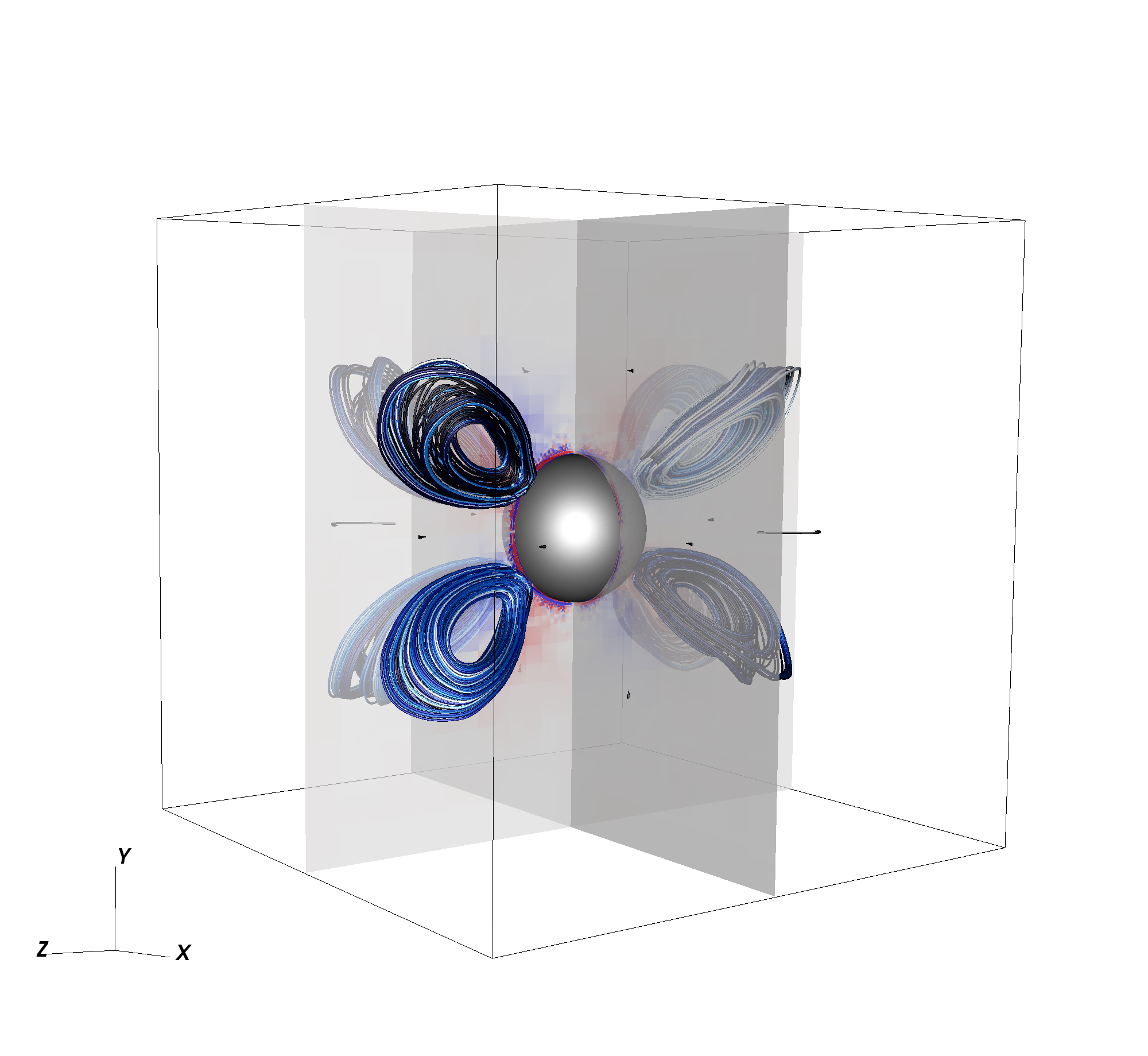}
			\subcaption{}
			\label{subfig:zr_he_t3}
			\end{subfigure}%
		%%%%%%%%%%%%%%%%%%%%%% 7 %%%%%%%%%%%%%%%%%%%%%%%%%%%%%%%%%%%
			\centering
			\begin{subfigure}{0.45\textwidth}	
			\centering
			\includegraphics[width=1.0\linewidth]{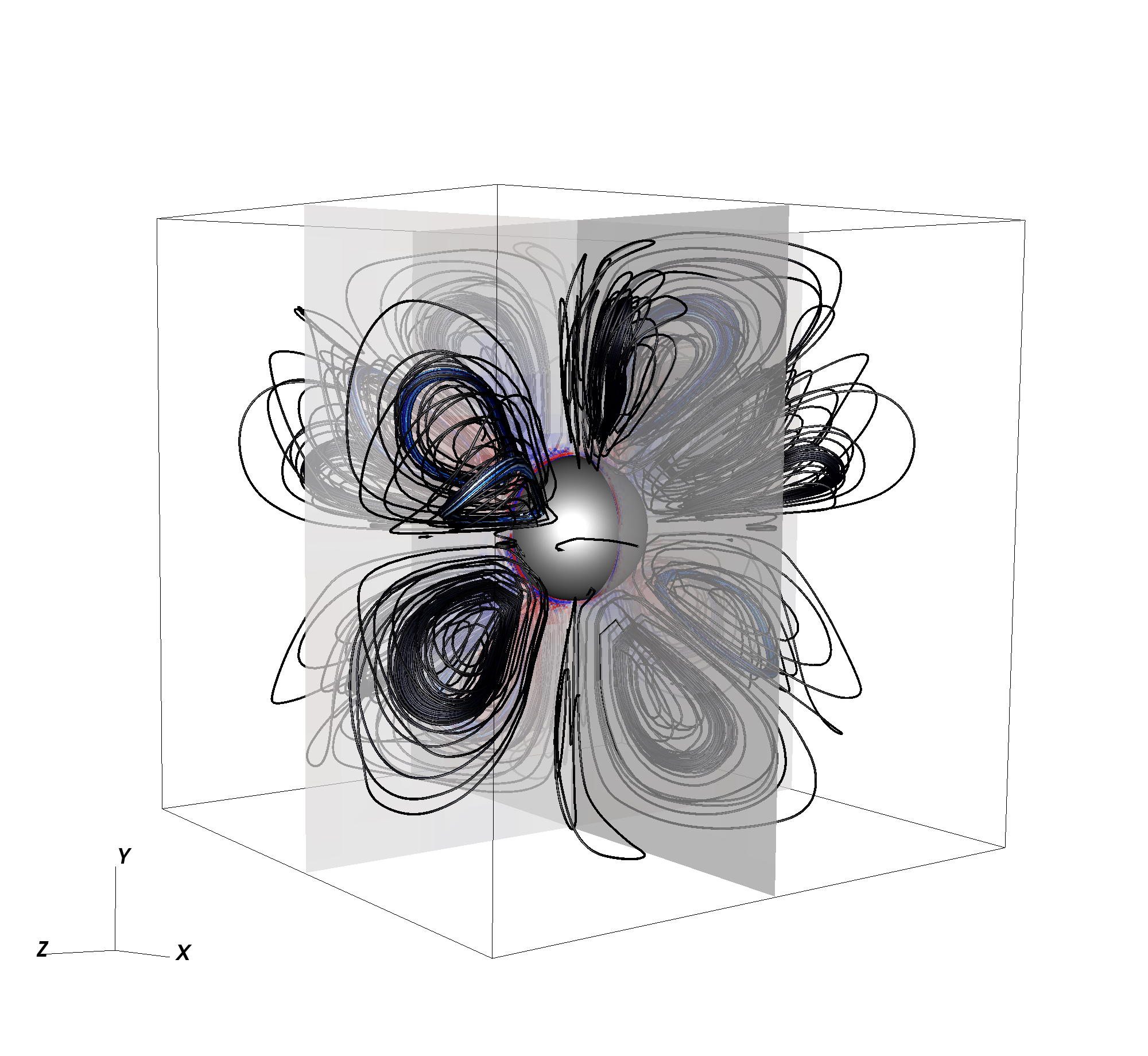}
			\subcaption{}
			\label{subfig:zr_he_t4}
			\end{subfigure}
   	\caption{\textit{Streamlines} of the Zirconium-Helium systems at different time values. The results were produced using $h=4.0/2^{10}$ to $h=4.0/2^4$, and $\delta t = \num{2e-5}$.}
	\label{fig:streamlinesZrHe}
%%%%%%%%%%%%%%%%%%%%%% 5 %%%%%%%%%%%%%%%%%%%%%%%%%%%%%%%%%%%
\end{figure}

To further understand the dynamics of the studied systems, we will show streamlines around the molten droplet in the Zirconium-Helium system in 3D. In~\cref{subfig:zr_he_timeshots}, we label 4 time values at which we will show streamlines. 

Plots in~\cref{fig:streamlinesZrHe} show the streamlines around the metal droplets at four time values. We note that several symmetric vortices are observed around the metal droplet, resulting from the surrounding fluid having a higher amount of kinetic energy manifesting itself into vortices. The reason for these large vortices is the relatively small (compared to the metal droplet) specific density and viscosity of the surrounding fluid trying to push back against the oscillating droplet. The demonstrated streamlines allow us to understand the damped oscillation of the droplet in terms of the evolution of these vortical flow structures in the surrounding fluids (Helium). The vortices around the droplet were smallest in size at $t_1$ as they were starting to develop (early in the droplet oscillation), as shown in ~\cref{subfig:zr_he_t1} where streamlines are still far away from the domain boundaries and near the Zirconium droplet. Upon further analysis, we note that the vortices at $t_3$ are also slightly smaller than those at the other time values ($t_2$ and $t_4$). This observation is because, at $t_3$, the droplet is at its equilibrium position after one full oscillation cycle, meaning that the interface velocity is close to zero. The number of these vortices and the time needed to transition from one set of vortices to another is a characteristic of the physical system density/viscosity ratio and interfacial tension.
The shown streamlines also explain the need to use a $[0,4] \times [0,4]$ box to simulate the oscillation dynamics in our physical systems, as streamlines extend close to the domain boundaries. All of the vortices shown in~\cref{fig:streamlinesZrHe} initiate in the Helium gas near the droplet interface. This explains the need to use a relatively fine mesh in the box surrounding the droplet's initial position to make the problem more stable and the linear solver less ill-conditioned.

\subsection{Physical Oscillations}
We plot the droplet shape oscillations for the three physical systems discussed in this paper. \cref{subfig:different_systems_2D} shows the shape oscillations in 2D while \cref{subfig:different_systems_3D} shows the shape oscillations in 3D. Using these plots, the physical frequency of oscillations is computed before calculating the interfacial tension using Rayleigh theory (\cref{eqn:rayleigh_ellipse_3D} and \cref{eqn:rayleigh_ellipse_2D}). For the 2D plots, we used a fine mesh with $h=4.0/2^{11}$ corresponding to 32 elements in the interface and $h=4.0/2^{8}$ away from the interface (background mesh). However, since the 3D cases are much more expensive, we used the coarsest mesh that provides stable results. The mesh used in 3D differs for each system as the interfacial tension, density, and viscosity ratio are different. For example, the less extreme density ratio cases such as Zirconium-Helium (1:36K) and Copper-Air (1:6K) simulations converge using a mesh with $h=4.0/2^{10}$ in the interface and $h=4.0/2^{4}$ away from the interface. On the other hand, the more extreme density ratio case of Osmium-Helium (1:117K) simulations requires a finer background mesh away from the interface with $h=4.0/2^{8}$. 
Due to the computational intensity of the 3D cases, particularly the Osmium-Helium scenario, which involves 20 million mesh nodes, we limit the 3D simulations to two complete oscillation cycles.

%Since the Osmium-Helium case is more computationally expensive with 20 million mesh nodes, we only run the simulations for one full oscillation cycle rather than two cycles for the other two cheaper cases.

\begin{figure} %[H]	
			\begin{subfigure}{\linewidth}	
				\centering
				\begin{tikzpicture}
				\begin{axis} [width=0.5\linewidth, xlabel={$t(s)$},ylabel={$A(t)$},legend columns=2,xmin=0, xmax=0.028,yticklabel style={/pgf/number format/precision=3}, legend style={at={(0.5,-0.2)},anchor=north}]
				\addplot [color=red]  table[x=time_ZrHe,
			    y=amp_ZrHe,
                col sep=comma,
                each nth point=10,
                filter discard warning=false] {Data/systems,Re=300,2D,physicalTime.csv};
				\addplot [color=ForestGreen]  table[x=time_CuAir,
				y=amp_CuAir,
				col sep=comma,
				each nth point=10,
				filter discard warning=false] {Data/systems,Re=300,2D,physicalTime.csv};
				\addplot [color=blue] table[x=time_OsHe,
				y=amp_OsHe,
				col sep=comma,
				each nth point=10,
				filter discard warning=false] {Data/systems,Re=300,2D,physicalTime.csv};   
				\legend{Zr-He,Cu-Air,Os-He}
				\end{axis}
				\end{tikzpicture} 
				%\centering
				\subcaption{2D Oscillations}
				\label{subfig:different_systems_2D}
				\end{subfigure} \\
				\begin{subfigure}{\linewidth}
				\centering
				\begin{tikzpicture}
				\begin{axis} [width=0.5\linewidth, xlabel={$t(s)$},ylabel={$A(t)$},legend columns=2,xmin=0, xmax=0.016,yticklabel style={/pgf/number format/precision=3}, legend style={at={(0.5,-0.2)},anchor=north}]
				\addplot [color=red]  table[x=time_ZrHe,
				y=amp_ZrHe,
                col sep=comma,
                each nth point=15,
                filter discard warning=false] {Data/systems_3D_Re=300.csv};
				\addplot [color=ForestGreen]  table[x=time_CuAir,
				y=amp_CuAir,
				col sep=comma,
				each nth point=15,
				filter discard warning=false] {Data/systems_3D_Re=300.csv};
				\addplot [color=blue]  table[x=time_OsHe,
				y=amp_OsHe,
				col sep=comma,
				each nth point=15,
				filter discard warning=false] {Data/systems_3D_Re=300.csv};
				\legend{Zr-He,Cu-Air, Os-He}
				\end{axis}
				\end{tikzpicture}
				\subcaption{3D Oscillations}
				\label{subfig:different_systems_3D}
				\end{subfigure}
	\caption{Panel a) 2D droplet oscillations in physical space for the three studied systems; panel b) 3D droplet oscillations in physical space for the three studied systems.}
	\label{fig:different_systems_response}
\end{figure}

\subsection{Energy Stability and Mass Conservation}
Even though the continuous CHNS model is thermodynamically consistent, not all spatial discretizations guarantee energy-stable solutions. Therefore, we must verify whether the fully discrete numerical implementation satisfies discrete energy stability and mass conservation. \Cref{subfig:energy_lvl10_dt1} and \cref{subfig:energy_lvl10_dt2} shows the evolution of $E_{tot}(\vec{v},\phi,t)$ as a function of $t(s)$ for two Zirconium-Helium cases using $h=4.0/2^{10}$ inside the interface and $h=4.0/2^4$ in the background mesh (two different time-steps). We can see from the inset of these plots that there is a small kink in the initial few time steps instead of an expected smooth decay of the energy functional. On the other hand, we also show in \cref{subfig:energy_lvl11_dt1} that $E_{tot}(\vec{v},\phi,t)$ decay smoothly without any instabilities or kinks when using a higher interfacial resolution with $h=4.0/2^{11}$ corresponding to 32 elements inside the interface. This empirically shows that increasing interfacial resolution can guarantee energy stability, especially for the high density/viscosity ratio cases. This is essentially due to large change in material properties (density and viscosity), which requires high interfacial resolution to resolve large pressure gradients across the interface.
Regarding mass conservation, we can see that all three spatial/temporal discretizations show similar behavior of excellent mass conservation for these simulations where the difference is extremely small and on the order of $10^{-4}$ (see \cref{fig:energyAndMassConsv}).

\begin{figure} %[htbp]
	\begin{subfigure}{0.45\textwidth}
		\centering
		\begin{tikzpicture}[spy using outlines={rectangle, magnification=2, size=0.5cm, connect spies}]
				\begin{axis} [width=0.99\linewidth, xlabel={$t(s)$},ylabel={$E_{tot}(\vec{v},\phi,t)$},xmin=0, xmax=0.016, scale=0.85]
						\addplot [color=red]  table[x=time,
						y=TotalEnergy,
						col sep=comma,
						each nth point=10,
						filter discard warning=false] {Data/EnergyData/Energy_data,lvl10,dt1.csv};
						\coordinate (a) at (axis cs:0.00025,34.656);
					\end{axis}
				\spy [black] on (a)in node  at (2.4,3.65);
			\end{tikzpicture}
		\subcaption{}
		\label{subfig:energy_lvl10_dt1}
	\end{subfigure}%
%%%
	\begin{subfigure}{0.45\textwidth}
	\centering
	\begin{tikzpicture}
				\begin{axis} [width=0.99\linewidth, xlabel={$t(s)$},ylabel={$\int_{\Omega} (\phi (\vec{x}) -\phi_{0}) \mathrm{d}\vec{x}$},xmin=0, xmax=0.016, ymin=-1e-4,ymax=1e4,ymin=-0.0001, ymax=0.0001, scale=0.85]
					\addplot [color=red]  table[x=time,
					y=MassDifference,
					col sep=comma,
					each nth point=10,
					filter discard warning=false] {Data/EnergyData/Energy_data,lvl10,dt1.csv};
					\end{axis}
				\end{tikzpicture}
	\subcaption{}
	\label{subfig:mass_lvl10_dt1}
	\end{subfigure}%
\\
	\begin{subfigure}{0.45\textwidth}
	\centering
    	\begin{tikzpicture}[spy using outlines={rectangle, magnification=2, size=0.5cm, connect spies}]
			\begin{axis} [width=0.99\linewidth, xlabel={$t(s)$},ylabel={$E_{tot}(\vec{v},\phi,t)$},xmin=0, xmax=0.016, scale=0.85]
				\addplot [color=red]  table[x=time,
				y=TotalEnergy,
				col sep=comma,
				each nth point=10,
				filter discard warning=false] {Data/EnergyData/Energy_data,lvl10,dt2.csv};
				\coordinate (a) at (axis cs:0.00025,34.656);
				\end{axis}
			\spy [black] on (a)in node  at (2.4,3.65);
			\end{tikzpicture}
	\subcaption{}
	\label{subfig:energy_lvl10_dt2}
	\end{subfigure}%
%%%
	\begin{subfigure}{0.45\textwidth}
	\centering
		\begin{tikzpicture}
			\begin{axis} [width=0.99\linewidth, xlabel={$t(s)$},ylabel={$\int_{\Omega} (\phi (\vec{x}) -\phi_{0}) \mathrm{d}\vec{x}$},xmin=0, xmax=0.016,ymin=-1e-4,ymax=1e4, ymin=-0.0001, ymax=0.0001,scale=0.85]
				\addplot [color=red]  table[x=time,
				y=MassDifference,
				col sep=comma,
				each nth point=10,
				filter discard warning=false] {Data/EnergyData/Energy_data,lvl10,dt2.csv};
				\end{axis}
			\end{tikzpicture}
	\subcaption{}
	\label{subfig:mass_lvl10_dt2}
	\end{subfigure}%
\\
	\begin{subfigure}{0.45\textwidth}
	\centering
	\begin{tikzpicture}[spy using outlines={rectangle, magnification=2, size=0.5cm, connect spies}]
		\begin{axis} [width=0.99\linewidth, xlabel={$t(s)$},ylabel={$E_{tot}(\vec{v},\phi,t)$},xmin=0, xmax=0.016, scale=0.85]
			\addplot [color=red]  table[x=time,
			y=TotalEnergy,
			col sep=comma,
			each nth point=10,
			filter discard warning=false] {Data/EnergyData/Energy_data,lvl11,dt1.csv};
			\coordinate (a) at (axis cs:0.00025,34.595);
			\end{axis}
		\spy [black] on (a)in node  at (2.4,3.65);
		\end{tikzpicture} 
	\subcaption{}
	\label{subfig:energy_lvl11_dt1}
	\end{subfigure}%
%%%
	\begin{subfigure}{0.45\textwidth}
	\centering
		\begin{tikzpicture}
			\begin{axis} [width=0.99\linewidth, xlabel={$t(s)$},ylabel={$\int_{\Omega} (\phi (\vec{x}) -\phi_{0}) \mathrm{d}\vec{x}$},xmin=0, xmax=0.016, ymin=-0.0001, ymax=0.0001, scale=0.85]
				\addplot [color=red]  table[x=time,
				y=MassDifference,
				col sep=comma,
				each nth point=10,
				filter discard warning=false] {Data/EnergyData/Energy_data,lvl11,dt1.csv};
				\end{axis}
			\end{tikzpicture}
	\subcaption{}
	\label{subfig:mass_lvl11_dt1}
	\end{subfigure}%
		\caption{Panel a) shows the decay of the total energy, with an inset zooming in on the initial kink using $h=4.0/2^{10}$ to $h=4.0/2^4$, and $\delta t=\num{1e-5}$; panel b) shows the corresponding mass conservation; panel c) shows the decay of the total energy, with an inset zooming in on the initial kink using $h=4.0/2^{10}$ to $h=4.0/2^4$, and $\delta t=\num{2e-5}$; panel d) shows the corresponding mass conservation; panel e) shows the (smooth) decay of the total energy using $h=4.0/2^{11}$ to $h=4.0/2^4$, and $\delta t=\num{1e-5}$; panel f) shows the corresponding mass conservation.}
	\label{fig:energyAndMassConsv}
\end{figure}

\subsection{Solver Configuration and Performance}
The linear systems we handle are fairly ill-conditioned due to the large density ratio and surface tension, causing high pressure gradients across the interface. For this reason, we find the stabilized biconjugate gradient linear solver with ASM preconditioner suitable for our problem. The command line options we provide~\petsc are in~\ref{petscCommands} for better reproduction. 
To monitor our solver performance, we report the average ksp iterations per time-step for the momentum and pressure-Poisson solver. We excluded the Cahn-Hilliard equation solver because it takes approximately one non-linear iteration to solve, with a fairly low number of linear solver ksp iterations. As seen in~\cref{tab:averageKSP}, the average number of ksp iterations for the pressure-Poisson solver is significantly larger than that of the momentum solver. The pressure-Poisson solver is harder to solve because the high surface tension and large density ratio cause a high pressure gradient across the interface.

\begin{table}[H] \centering
%\centering\scriptsize\setlength\tabcolsep{2pt}
\begin{tabular}{@{}|c|c|c|c|c|@{}}
\toprule
{System}	    &  {momentum} & {pressure-Poisson} & {time to solve (hr)} & {mesh nodes (million)} \\ 
\midrule
\midrule
{Osmium-Helium}    & {$3.1$}                                              & {$19.6$}                                           & {$486$}    & {$20$}                  \\
{Zirconium-Helium} & {$4.1$}                                            & {$23.4$}                                         & {$72$} & {$3.5$}                    \\
{Copper-Air}    & {$3.5$}                                              & {$23.4$}                                           & {$84$}    & {$3.5$}                  \\
\bottomrule
\end{tabular}
\caption{Comparison of the average number of ksp iterations per time-step for the momentum and pressure-Poisson solver, the average number of computing hours needed to \textbf{finish one full cycle}, and the total number of mesh nodes needed for the three systems in 3D.}
\label{tab:averageKSP} 
\end{table}

All simulations presented in this paper were performed on $1024$ cores using $8$ computing nodes with 128 cores each on~\bridges.

\subsection{Interfacial Tension}
 After obtaining convergent oscillation results, we calculate the oscillation frequency for the three systems under consideration using the fast Fourier transform. Further, we can then calculate the metals' surface tension based on the Rayleigh equation (\cref{eqn:rayleigh_ellipse_3D} and \cref{eqn:rayleigh_ellipse_2D}). ~\Cref{tab:surfaceTension} compares the calculated and experimental surface tension measurements for Osmium, Zirconium, and Copper droplets in both 2D and 3D. All calculated surface tensions are within a 5\% margin of error in both dimensions. 
 
 %These results explain the need for high interfacial resolution with extreme density change across a thin region. For the Osmium-Helium case in 3D, the error in interfacial tension is less than in the other two cases since we used a higher background mesh resolution $h=4.0/2^8$ as it is necessary to get linear solver convergence. It is important to note that surface tension is inherently 3D in nature. This means that validating our simulations with experiments makes 3D simulations necessary, which explains the necessity to simulate expensive 3D simulations even if we can compute the interfacial tension from 2D simulations. 
 Another key observation is that our calculated interfacial tension is generally less than the experimental value. This result can be explained by the fact that Rayleigh's theory assumes inviscid droplets while our simulation considers viscous forces.

\begin{table}[H] \centering
\begin{tabular}{@{}|c|c|c|c|c|@{}}
\toprule
{System} & {Experimental $\sigma$ (\si{\newton\per\meter})} & {$\sigma$ in 2D (\si{\newton\per\meter})} & {$\sigma$ in 3D (\si{\newton\per\meter})} \\
\midrule
\midrule
{Osmium} & {$2.7$} & {$2.65$} & {$2.58$} \\
{Zirconium} & {$1.45$} & {$1.42$} & {$1.36$} \\
{Copper} & {$1.16$} & {$1.18$} & {$1.08$} \\
\bottomrule
\end{tabular}
\caption{Comparison of experimental and computed surface tension values in 2D and 3D for the three physical systems analyzed \cite{Marcus2016, Sumaria2019, Sumaria2017}.}
\label{tab:surfaceTension}
\end{table}

\section{Conclusions}

In this study, we have successfully demonstrated the capability of a projection-based Cahn-Hilliard Navier-Stokes (CHNS) framework to model very high-density ratio ($10^4-10^5:1$) two-phase flows, specifically focusing on the dynamics of molten metal droplets in microgravity. Our approach leverages a thermodynamically consistent diffuse interface model, which has shown robust performance in handling the large density and viscosity ratios characteristic of these systems. To further validate the robustness and versatility of our framework, we simulated several canonical problems, including the single bubble rise and capillary wave problems, across a range of density ratios.

We have addressed several key challenges of modeling such extreme conditions through extensive numerical simulations. First, we quantified the impact of mesh resolution on energy stability and mass conservation, and our results show that higher interfacial mesh resolution is crucial for maintaining energy stability and avoiding numerical instabilities. This finding is particularly important for accurately capturing the oscillation dynamics of the droplets. Second, the use of adaptive mesh refinement proved to be essential in efficiently resolving the interface while maintaining computational efficiency. Our scalable octree-based adaptive mesh approach enabled precise simulation of complex physical phenomena in 3D by effectively tracking highly deforming interfaces. Furthermore, validation of our numerical results, particularly regarding the calculated interfacial tensions, showed agreement within a 5\% margin of error, highlighting the accuracy of our model in replicating real-world behaviors. This combination of advanced modeling and validation establishes the reliability of the CHNS framework for practical applications in material science and manufacturing.

In conclusion, the projection-based CHNS framework offers a powerful and efficient tool for modeling complex two-phase flow systems. By addressing the critical issues of mesh resolution and computational efficiency, our work sets the stage for more accurate and scalable simulations, contributing to the broader fluid dynamics and materials science field. Future research directions that build on this work include (a) extension to more complex scenarios like three-phase flows -- for instance, slag/melt/air systems, which are critical for continuous casting and other manufacturing applications; (b) extension to include electro-/magneto-hydrodynamic effects for modeling advanced manufacturing of critical materials, and (c) deploying this framework across more complex geometry applications.

%This paper deployed a projection-based second-order in-time numerical framework to solve the Cahn-Hilliard Navier-Stokes (CHNS) equations for two-phase flows. We used a conforming Galerkin method for spatial discretization using linear basis functions with VMS stabilization. A simple modification to the VMS approach is proposed to make the model more stable for simulating high density and viscosity ratio two-phase flows. Numerical studies showed droplet shape oscillation convergence with mesh resolution, timestep, and interfacial thickness. We have also numerically verified that the scheme is energy-stable and mass-conserving at the discrete level for the Zirconium-Helium system. Due to our simulations' high surface tension and density ratio, a fine spatial resolution is required to resolve the interface in 3D accurately. This projection-based framework can be extended to 3D with adaptive mesh refinement necessary to resolve the interface of these extreme density and viscosity ratio systems ($10^4-10^5:1$). 

%We also showed that the projection-based framework is more computationally efficient, even if it requires a smaller timestep for the convergence of the linear system. 
\section{Acknowledgements}
The authors gratefully acknowledge funding from the National Science Foundation 2053760 and NASA under grant 80NSSC20K1038. The authors also acknowledge XSEDE grant number TG-CTS110007 for computing time on TACC~\stampede, ALCF computing resources, and Iowa State University computing resources. This work used the Extreme Science and Engineering Discovery Environment (XSEDE), supported by National Science Foundation grant number ACI-1548562. Specifically, it used the Bridges system, supported by NSF award number ACI-1445606, at the Pittsburgh Supercomputing Center (PSC).
\appendix
\section{PETSc Command Options}
\label{petscCommands}
For better reproduction, we provide the PETSc command line options, which we used to simulate our cases. Upon trying different solvers, we found that the biCGstab-based Krylov solver with Additive Schwarz preconditioning is the most efficient for solving the 4 equations. $ksp\_atol$ and $ksp\_rtol$ in the commands below are the absolute and the relative tolerances of the linear solver, respectively. Similarly, $snes\_atol$ and $snes\_rtol$ in the commands below are the absolute and the relative tolerances of the nonlinear solver used for the Cahn-Hilliard block solver, respectively. Some of the below commands are used to print norms and monitor the solver performance. 

\begin{lstlisting}
solver_options_momentum = {
  ksp_atol = 1e-8
  ksp_rtol = 1e-7
  ksp_max_it = 2000
  ksp_type = "bcgs"
  pc_type = "asm"
  ksp_monitor = ""
  ksp_converged_reason = ""
};
solver_options_pp = {
  ksp_atol = 1e-8
  ksp_rtol = 1e-7
  ksp_max_it = 2000
  ksp_type = "bcgs"
  pc_type = "asm"
  ksp_monitor = ""
  ksp_converged_reason = ""
};
solver_options_vupdate = {
  ksp_atol = 1e-8
  ksp_rtol = 1e-7
  ksp_max_it = 2000
  ksp_type = "bcgs"
  pc_type = "asm"
  ksp_monitor = ""
  ksp_converged_reason = ""
};
solver_options_ch = {
  snes_rtol = 1e-10
  snes_atol = 1e-10
  ksp_atol = 1e-8
  ksp_rtol = 1e-7
  ksp_max_it = 2000
  ksp_type = "bcgs"
  pc_type = "asm"
  ksp_monitor = ""
  ksp_converged_reason = ""
};
\end{lstlisting}

\bibliographystyle{elsarticle-num-names}
\bibliography{total_references}

\end{document}